\RequirePackage{ifpdf}
\ifpdf 
\documentclass[pdftex]{sigma}
\else
\documentclass{sigma}
\fi

\numberwithin{equation}{section}

\usepackage{local_qg}

\def\cstok#1{\leavevmode\thinspace\hbox{\vrule\vtop{\vbox{\hrule\kern1pt
\hbox{\vphantom{\tt/}\thinspace{\tt#1}\thinspace}}
\kern1pt\hrule}\vrule}\thinspace}

\begin{document}

\allowdisplaybreaks

\renewcommand{\PaperNumber}{098}

\FirstPageHeading

\renewcommand{\thefootnote}{$\star$}

\ShortArticleName{Quantum Gravity: Unif\/ication of Principles,
Promises of Spectral Geometry}

\ArticleName{Quantum Gravity: Unif\/ication of Principles\\ and Interactions,
and Promises of Spectral Geometry\footnote{This paper is a
contribution to the Proceedings of the 2007 Midwest
Geometry Conference in honor of Thomas~P.\ Branson. The full collection is available at
\href{http://www.emis.de/journals/SIGMA/MGC2007.html}{http://www.emis.de/journals/SIGMA/MGC2007.html}}}

\Author{Bernhelm BOOSS-BAVNBEK~$^\dag$, Giampiero ESPOSITO~$^\ddag$
and Matthias LESCH~$^\S$}

\AuthorNameForHeading{B.~Boo{\ss}-Bavnbek, G.~Esposito and M.~Lesch}

\Address{$^\dag$~IMFUFA, Roskilde University, P.O. Box 260,
4000 Roskilde, Denmark}

\EmailD{\href{mailto:booss@ruc.dk}{booss@ruc.dk}}
\URLaddressD{\url{http://imfufa.ruc.dk/~Booss}}

\Address{$^\ddag$~INFN, Sezione di Napoli and Dipartimento di Scienze
Fisiche, \\
$\phantom{^\ddag}$~Complesso Universitario di Monte S. Angelo, Via Cintia,
Edif\/icio 6, 80126 Napoli, Italy}

\EmailD{\href{mailto:giampiero.esposito@na.infn.it}{giampiero.esposito@na.infn.it}}
\URLaddressD{\url{http://www.dsf.unina.it/Theor/gruppoiv/gesposit.htm}}

\Address{$^\S$~Bonn University, Mathematical Institute,
Beringstr. 6, D-53115 Bonn, Germany}
\EmailD{\href{mailto:lesch@math.uni-bonn.de}{lesch@math.uni-bonn.de}}

\URLaddressD{\url{http://www.math.uni-bonn.de/people/lesch}}

\ArticleDates{Received August 14, 2007, in f\/inal form September 25, 2007; Published online October 05, 2007}

\Abstract{Quantum gravity was born as that branch of modern
theoretical physics that tries to unify its guiding principles,
i.e., quantum mechanics and general relativity. Nowadays it is
providing new insight into the unif\/ication of all fundamental
interactions, while giving rise to new developments in modern
mathematics. It is however unclear whether it will ever become a
falsif\/iable physical theory, since it deals with Planck-scale
physics. Reviewing a~wide range of spectral geometry from index
theory to spectral triples, we hope to dismiss the general opinion
that the mere mathematical complexity of the unif\/ication programme
will obstruct that programme.}


\Classification{83C45; 83E05; 83E30; 83E50; 58B34; 58J20; 58J28;
58J30; 58J32; 58J52}

\rightline{\it In memory of Tom Branson}

\renewcommand{\thefootnote}{\arabic{footnote}}
\setcounter{footnote}{0}

\section{Introduction}

\subsection*{Lorentzian spacetime and gravity}
In modern physics, thanks to the work of Einstein \cite{Eins16},
space and time are unif\/ied into the spacetime manifold $(M,g)$, where
the metric $g$ is a real-valued symmetric bilinear map
\[
g:T_{p}(M) \times T_{p}(M) \rightarrow \R
\]
of Lorentzian signature. The latter feature gives rise to the light-cone
structure of spacetime, with vectors being divided into timelike, null or
spacelike depending on whether $g(X,X)$ is negative, vanishing or positive,
respectively. The classical laws of nature are written in tensor language,
and gravity is the curvature of spacetime. In the theory of general
relativity, gravity couples to the energy-momentum tensor of matter through
the Einstein equations
\begin{equation}
R_{ab}-\frac{1}{2}g_{ab}R={8 \pi G \over c^{4}}T_{ab}.
\label{(1)}
\end{equation}
The Einstein--Hilbert action functional for gravity, giving rise to
equation~\eqref{(1)}, is dif\/feo\-mor\-phism-invariant, and hence general relativity
belongs actually to the general set of theories ruled by an
inf\/inite-dimensional~\cite{DeWi65} invariance group
(or pseudo-group). With hindsight, following DeWitt~\cite{DeWi05},
one can say that general relativity was actually the
f\/irst example of a non-Abelian gauge theory (about 38 years before
Yang--Mills theory~\cite{YaMi54}).

\subsection*{From Schr\"{o}dinger to Feynman}

Quantum mechanics deals instead, mainly, with a description of
the world on atomic or sub-atomic scale. It tells us that, on such
scales, the world can be described by a Hilbert space structure, or
suitable generalizations. Even in the relatively simple case of the
hydrogen atom, the appropriate Hilbert space is inf\/inite-dimensional, but
f\/inite-dimensional Hilbert spaces play a role as well. For example, the
space of spin-states of a spin-$s$ particle is $\C^{2s+1}$ and is therefore
f\/inite-dimensional. Various pictures or formulations of quantum mechanics
have been developed over the years, and their key elements can be summarized
as follows:
\renewcommand{\labelenumi}{\textup{(\roman{enumi})}}
\begin{enumerate}
\itemsep=0pt

\item In the {\it Schr\"{o}dinger picture}, one deals with wave functions
evolving in time according to a f\/irst-order equation. More precisely,
in an abstract Hilbert space ${\cal H}$, one studies the Schr\"{o}dinger
equation
\begin{gather*}
\textup{i}{\hbar}{d \psi \over dt}={\hat H} \psi,
\end{gather*}
where the state vector $\psi$ belongs to ${\cal H}$, while ${\hat H}$ is
the Hamiltonian operator. In wave mechanics, the emphasis is more
immediately put on partial dif\/ferential equations, with the wave function
viewed as a complex-valued map $\psi: (x,t) \rightarrow \C$ obeying
the equation
\begin{gather*}
\textup{i}{\hbar}{\partial \psi \over \partial t}
=\left(-{{\hbar}^{2}\over 2m}\bigtriangleup + V \right) \psi,
\end{gather*}
where $- \bigtriangleup$ is the Laplacian in Cartesian coordinates on
$\R^{3}$ (with this sign convention, its symbol is positive-def\/inite).
\item In the {\it Heisenberg picture}, what evolves in time are instead
the operators, according to the f\/irst-order equation
\begin{gather*} \textup{i}{\hbar}{d{\hat A}\over dt}=[{\hat A},{\hat H}].
\end{gather*}
Heisenberg performed a quantum mechanical re-interpretation of kinematic
and mechanical relations \cite{Heis25}
because he wanted to formulate quantum theory
in terms of observables only.

\item  In the {\it Dirac quantization}, from an assessment of the
Heisenberg approach and of Poisson brackets \cite{Dira25},
one discovers that quantum
mechanics can be made to rely upon the basic commutation relations
involving position and momentum operators:
\begin{gather}
[{\hat q}^{j},{\hat q}^{k}]=[{\hat p}_{j},{\hat p}_{k}]=0,
\label{(5)}
\\
[{\hat q}^{j},{\hat p}_{k}]=\textup{i}{\hbar}\delta_{\; k}^{j}.
\label{(6)}
\end{gather}
For generic operators depending on ${\hat q}$, ${\hat p}$ variables,
their formal Taylor series, jointly with application of \eqref{(5)} and
\eqref{(6)}, should yield their commutator.
\item  {\it Weyl quantization}. The operators satisfying the canonical
commutation relations \eqref{(6)} cannot be both bounded, whereas it would be nice
to have quantization rules not involving unbounded operators and domain
problems. For this purpose, one can consider the strongly continuous
one-parameter unitary groups having position and momentum as their
inf\/initesimal generators. These read as $U(s) \equiv e^{\textup{i}sP}$,
$V(t) \equiv e^{\textup{i}tQ}$, and satisfy the Weyl form of canonical
commutation relations, which is given by
\begin{gather*}
U(s)V(t)=e^{\textup{i}st}V(t)U(s).
\end{gather*}
Here the emphasis was, for the f\/irst time, on group-theoretical methods,
with a substantial departure from the historical development, that relied
instead heavily on quantum commutators and their relation with
classical Poisson brackets.
\item {\it Feynman quantization} (i.e., Lagrangian approach).
The Weyl approach is very elegant and far-sighted,
with several modern applications
\cite{Espo04}, but still has to do with a more rigorous
way of doing canonical quantization, which is not suitable for an
inclusion of relativity. A spacetime approach to ordinary quantum mechanics
was instead devised by Feynman \cite{Feyn48}
(and partly Dirac himself \cite{Dira33}), who proposed
to express the Green kernel of the Schr\"{o}dinger equation in the form
\begin{gather*}
G[x_{f},t_{f};x_{i},t_{i}]=\int_{{\rm all \; paths}}
e^{\textup{i}S / {\hbar}} d\mu,
\end{gather*}
where $d\mu$ is a suitable (putative) measure on the set of all space-time
paths (including continuous, piecewise continuous, or even discontinuous
paths) matching the initial and f\/inal conditions. This point of view has
enormous potentialities in the quantization of f\/ield theories, since it
preserves manifest covariance and the full symmetry group, being derived
from a Lagrangian.
\end{enumerate}

It should be stressed that quantum mechanics regards wave functions
only as a technical tool to study bound states (corresponding to the
discrete spectrum of the Hamiltonian operator~${\hat H}$),
scattering states (corresponding instead to the continuous spectrum
of ${\hat H}$), and to evaluate probabilities (of f\/inding the values
taken by the observables of the theory). Moreover, it is meaningless to
talk about an elementary phenomenon on atomic (or sub-atomic) scale
unless it is registered, and quantum mechanics in the laboratory
needs also an external observer and assumes the so-called reduction
of the wave packet (see \cite{Espo04} and references therein).
\subsection*{Spacetime singularities}
Now we revert to the geometric side. In Riemannian or pseudo-Riemannian
geometry, geodesics are curves whose tangent vector $X$ moves by
parallel transport, so that eventually
\begin{gather*}
{dX^{a}\over ds}+\Gamma_{\; bc}^{a}X^{b}X^{c}=0,
\end{gather*}
where $s$ is the af\/f\/ine parameter and $\Gamma_{\; bc}^{a}$ are the
connection coef\/f\/icients. In general relativi\-ty, timelike geodesics
correspond to the trajectories of freely moving observers, while
null geodesics describe the trajectories of zero-rest-mass particles
\cite[Section~8.1]{HaEl73}. Moreover, a~spacetime~$(M,g)$ is said to be singularity-free if {\it all timelike
and null geodesics can be extended to arbitrary values of their
affine parameter}. At a spacetime singularity in general relativity,
all laws of classical physics would break down, because one would witness
very pathological events such as the sudden disappearance of freely
moving observers, and one would be completely unable to predict what
came out of the singularity.
In the sixties, Penrose \cite{Penr65}
proved f\/irst an important theorem on the
occurrence of singularities in gravitational collapse (e.g.\ formation
of black holes). Subsequent work by Hawking
\cite{Hawk65, Hawk66a, Hawk66b, Hawk66c,Hawk67},
Geroch \cite{Gero66}, Ellis and Hawking \cite{HaEl65, ElHa68},
Hawking and Penrose \cite{HaPe70}
proved that spacetime singularities are generic
properties of general relativity, provided that physically realistic
energy conditions hold. Very little analytic use of the Einstein
equations is made, whereas the key role emerges of topological and
global methods in general relativity (jointly with Morse theory for
Lorentzian manifolds).

\subsection*{Unif\/ication of all fundamental interactions}
The fully established unif\/ications of modern physics are as follows.
\begin{enumerate}
\itemsep=0pt

\item {\it Maxwell}: electricity and magnetism are unif\/ied into
electromagnetism. All related phenomena can be described by an
antisymmetric rank-two tensor f\/ield, and derived from a~one-form,
called the potential.
\item {\it Einstein}: space and time are unif\/ied into the spacetime
manifold. Moreover, inertial and gravitational mass, conceptually dif\/ferent,
are actually unif\/ied as well.
\item {\it Standard model of particle physics}: electromagnetic, weak
and strong forces are unif\/ied by a non-Abelian gauge theory, normally
considered in Minkowski spacetime (this being the base space in
f\/ibre-bundle language).
\end{enumerate}

The physics community is now familiar with a picture relying upon four
fundamental interactions: electromagnetic, weak, strong and gravitational.
The large-scale structure of the universe, however, is ruled by gravity
only. All unif\/ications beyond Maxwell involve non-Abelian gauge groups
(either Yang--Mills or dif\/feomorphism group). Three extreme views have
been developed along the years, i.e.,
\begin{enumerate}
\itemsep=0pt

\item  Gravity arose f\/irst, temporally,
in the very early Universe, then all other fundamental interactions.
\item  Gravity might result from Quantum Field Theory (this was the
Sakharov idea \cite{Sakh68}).
\item  The vacuum of particle physics is regarded as a cold quantum liquid
in equilibrium. Protons, gravitons and gluons are viewed as collective
excitations of this liquid \cite{Volo03}.
\end{enumerate}

\section[From old to new unification]{From old to new unif\/ication}\label{Section2}

Here we outline how the space-of-histories formulation provides a common ground for describing the `old' and `new' unif\/ications of fundamental theories.
\subsection*{Old unif\/ication}
Quantum f\/ield theory begins once an action functional $S$ is given, since
the f\/irst and most fundamental assumption of quantum theory is that every
isolated dynamical system can be described by a characteristic action
functional \cite{DeWi65}.  The Feynman approach makes it necessary to
consider an inf\/inite-dimensional manifold such as the space $\Phi$ of all
f\/ield histories $\varphi^{i}$.
On this space there exist (in the case of gauge theories) vector f\/ields
\begin{gather*}
Q_{\alpha}=Q_{\; \alpha}^{i} {\delta \over \delta \varphi^{i}}
\end{gather*}
that leave the action invariant, i.e., \cite{DeWi05}
\begin{gather}
Q_{\alpha}S=0.
\label{(11)}
\end{gather}
The Lie brackets of these vector f\/ields lead to a classif\/ication
of all gauge theories known so far.
\subsection*{Type-I gauge theories}
The peculiar property of type-I gauge theories is that these Lie
brackets are equal to linear combinations of the vector f\/ields themselves,
with structure constants, i.e., \cite{DeWi03}
\begin{gather}
[Q_{\alpha},Q_{\beta}]=C_{\; \alpha \beta}^{\gamma} \; Q_{\gamma},
\label{(12)}
\end{gather}
where $C_{\; \alpha \beta ,i}^{\gamma}=0$. The Maxwell, Yang--Mills,
Einstein theories are all examples of type-I theories (this is the
`unifying feature'). All of them, being gauge theories, need
supplementary conditions, since the second functional derivative of
$S$ is not an invertible operator. After imposing such conditions, the
theories are ruled by a dif\/ferential operator of d'Alembert type (or
Laplace type, if one deals instead with Euclidean f\/ield theory), or a
non-minimal operator at the very worst (for arbitrary choices of
gauge parameters). For example, when Maxwell theory is quantized via
functional integrals in the Lorenz \cite{Lore67} gauge, one deals
with a gauge-f\/ixing functional
\begin{gather*}
\Phi(A)=\nabla^{b}A_{b},
\end{gather*}
and the second-order dif\/ferential operator acting on the potential
reads as
\begin{gather*}
P_{a}^{\; b}=-\delta_{a}^{\; b}\cstok{\ }+R_{a}^{\; \; b}
+\left(1-{1\over \alpha}\right)\nabla_{a}\nabla^{b},
\end{gather*}
where $\alpha$ is an arbitrary gauge parameter. The Feynman choice
$\alpha=1$ leads to the minimal operator
\[
{\widetilde P}_{a}^{\; b}=-\delta_{a}^{\; b}\cstok{\ }+R_{a}^{\; b},
\]
which is the standard wave operator on vectors in curved spacetime.
Such operators play a~leading role in the one-loop expansion of the
Euclidean ef\/fective action, i.e., the quadratic order in $\hbar$ in
the asymptotic expansion of the functional ruling the quantum theory
with positive-def\/inite metrics (see Section~\ref{Section6}).

\subsection*{Type-II gauge theories}
For type-II gauge theories, Lie brackets of vector f\/ields $Q_{\alpha}$ are
as in equation~\eqref{(12)} for type-I theories, but the structure constants are promoted
to structure functions. An example is given by simple supergravity (a
supersymmetric \cite{GoLi71,WeZu74} gauge theory of gravity, with a symmetry
relating bosonic and fermionic f\/ields) in four spacetime dimensions, with
auxiliary f\/ields \cite{Nieu81}.

\subsection*{Type-III gauge theories}
In this case, the Lie bracket \eqref{(12)} is generalized by
\begin{gather*}
[Q_{\alpha},Q_{\beta}]=C_{\; \alpha \beta}^{\gamma} \; Q_{\gamma}
+U_{\; \alpha \beta}^{i} \; S_{,i},
\end{gather*}
and it therefore reduces to \eqref{(12)} only on the {\it mass-shell}, i.e.,
for those f\/ield conf\/igurations satisfying the Euler--Lagrange equations.
An example is given by theories with gravitons and gravitinos such as
Bose--Fermi supermultiplets of both simple and extended supergravity
in any number of spacetime dimensions, without auxiliary
f\/ields \cite{Nieu81}.

\subsection*{From supergravity to general relativity}
It should be stressed that general relativity is naturally related to
supersymmetry, since the requirement of gauge-invariant Rarita--Schwinger
equations \cite{RaSc41} implies Ricci-f\/latness in four dimensions
\cite{DeZu76}, which is then equivalent to vacuum Einstein equations. The
Dirac operator \cite{Espo98} is more fundamental in this framework, since
the $m$-dimensional spacetime metric is entirely re-constructed from the
$\gamma$-matrices, in that \begin{gather*} g^{ab}={1\over 2m} \tr
(\gamma^{a}\gamma^{b}+\gamma^{b}\gamma^{a}).
\end{gather*}

\subsection*{New unif\/ication}
In modern high energy physics, the emphasis is no longer on f\/ields (sections
of vector bundles in classical f\/ield theory, operator-valued distributions in
quantum f\/ield theory), but rather on extended objects such as strings
\cite{Deli99}. In string theory, particles are not described as points, but
instead as strings, i.e., one-dimensional extended objects. While a point
particle sweeps out a one-dimensional worldline, the string sweeps out a
worldsheet, i.e., a two-dimensional real surface. For a free string, the
topology of the worldsheet is a cylinder in the case of a closed string, or a
sheet for an open string. It is assumed that dif\/ferent elementary particles
correspond to dif\/ferent vibration modes of the string, in much the same way
as dif\/ferent minimal notes correspond to dif\/ferent vibrational modes of
musical string instruments \cite{Deli99}.
The f\/ive dif\/ferent string theories \cite{AnSa02} are
dif\/ferent aspects of a more fundamental theory, called
$M$-theory \cite{Beck07}.
In the latest developments, one deals with `branes', which are classical
solutions of the equations of motion of the low-energy string ef\/fective
action, that correspond to new non-perturbative states of string theory,
break half of the supersymmetry, and are required by T-duality in theories
with open strings. They have the peculiar property that open strings have
their end-points attached to them \cite{DiLi99a, DiLi99b}.  With the language
of pseudo-Riemannian geometry, branes are timelike surfaces embedded into
bulk spacetime \cite{BaNe06,Barv06}. According to this picture, gravity
lives on the bulk, while standard-model gauge f\/ields are conf\/ined on the
brane. For branes, the normal vector $N$ is spacelike with respect to the
bulk metric $G_{AB}$, i.e.,
\begin{gather*}
G_{AB}N^{A}N^{B}=N_{C}N^{C} >0.
\end{gather*}
The action functional $S$ splits into the sum \cite{Barv06}
($g_{\alpha \beta}(x)$ being the brane metric)
\begin{gather*}
S=S_{4}[g_{\alpha \beta}(x)]+S_{5}[G_{AB}(X)],
\end{gather*}
while the ef\/fective action \cite{DeWi03} $\Gamma$ is formally given by
\begin{gather*}
e^{\textup{i}\Gamma}=\int DG_{AB}(X) \; e^{\textup{i}S} \times
\text{g.f. term}.
\end{gather*}
In the functional integral, the gauge-f\/ixed action reads as
(here there is summation as well as integration over repeated indices
\cite{DeWi65, DeWi03, Barv06})
\begin{gather*}
S_{{\rm g.f.}}=S_{4}+S_{5}+\tfrac{1}{2}F^{A}\Omega_{AB}F^{B}
+\tfrac{1}{2}\chi^{\mu}\omega_{\mu \nu}\chi^{\nu},
\end{gather*}
where $F^{A}$ and $\chi^{\mu}$ are bulk and brane gauge-f\/ixing functionals,
respectively, while $\Omega_{AB}$ and $\omega_{\mu \nu}$ are non-singular
`matrices' of gauge parameters.
The gauge-invariance properties of bulk and brane action functionals
can be expressed by saying that there exist vector f\/ields on the
space of histories such that (cf.\ equation~\eqref{(11)})
\begin{gather*}
R_{B}S_{5}=0, \qquad R_{\nu}S_{4}=0,
\end{gather*}
whose Lie brackets obey a relation formally analogous to equation~\eqref{(12)}
for ordinary type-I theories, i.e.,
\begin{gather*}
[R_{B},R_{D}]=C_{\; BD}^{A} \; R_{A}, \qquad
[R_{\mu},R_{\nu}]=C_{\; \mu \nu}^{\lambda} \; R_{\lambda}.
\end{gather*}
The bulk and brane ghost operators (see \cite{DeWi67, FaPo67} for the f\/irst
time that ghost f\/ields were considered in quantized gauge theories. When
one studies how the gauge-f\/ixing functional behaves under inf\/initesimal
gauge transformations, one discovers that one can then def\/ine a
dif\/ferential operator acting on f\/ields $\chi^{\beta}$ of the opposite
statistics with respect to the f\/ields occurring in the gauge-invariant
action functional. Such f\/ields $\chi^{\beta}$ are the ghost f\/ields.
They arise entirely from the f\/ibre-bundle structure of the space of
histories, from the Jacobian of the transformation from f\/ibre-adapted
coordinates to the conventional local f\/ields.) are therefore
\begin{gather*}
Q_{\; B}^{A}=R_{B}F^{A}=F_{\; ,a}^{A} \; R_{\; B}^{a},
\qquad
J_{\; \nu}^{\mu}=R_{\nu}\chi^{\mu}=\chi_{\; ,i}^{\mu} \;
R_{\; \nu}^{i},
\end{gather*}
respectively. The full bulk integration means integrating f\/irst with
respect to all bulk metrics~$G_{AB}$ inducing on the boundary
${\partial M}$ the given brane metric $g_{\alpha \beta}(x)$, and then
integrating with respect to all brane metrics.
Thus, one f\/irst evaluates the cosmological wave function of the bulk
spacetime \cite{Barv06}, i.e.,
\begin{gather*}
\psi_{\rm Bulk}=\int_{G_{AB}[\partial M]=g_{\alpha \beta}}
\mu(G_{AB},S_{C},T^{D})e^{\textup{i}{\widetilde S}_{5}},
\end{gather*}
where $\mu$ is taken to be a suitable measure, the $S_{C}$, $T^{D}$ are
ghost f\/ields, and
\begin{gather*}
{\widetilde S}_{5} \equiv S_{5}[G_{AB}]
+\tfrac{1}{2}F^{A}\Omega_{AB}F^{B}+S_{A}Q_{\; B}^{A} T^{B}.
\end{gather*}
Eventually, the ef\/fective action results from
\begin{gather*}
e^{\textup{i}\Gamma}=\int
{\widetilde \mu}(g_{\alpha \beta},\rho_{\gamma},\sigma^{\delta})
e^{\textup{i}{\widetilde S}_{4}}\psi_{\rm Bulk},
\end{gather*}
where ${\widetilde \mu}$ is another putative measure, $\rho_{\gamma}$ and
$\sigma^{\delta}$ are brane ghost f\/ields, and
\begin{gather*}
{\widetilde S}_{4} \equiv S_{4}+\tfrac{1}{2}\chi^{\mu}\omega_{\mu \nu}
\chi^{\nu}+\rho_{\mu}J_{\; \nu}^{\mu} \sigma^{\nu}.
\end{gather*}

\section{Open problems}\label{Section3}

Although string theory may provide a f\/inite theory of quantum gravity
that unif\/ies all fundamental interactions at once, its impact on particle
physics phenomenology and laboratory experiments remains elusive. Some key
issues are therefore in sight:
\begin{enumerate}
\itemsep=0pt

\item  What is the impact (if any) of Planck-scale physics on cosmological
observations \cite{COBEWMAP}?
\item  Will general relativity retain its role of fundamental theory, or
shall we have to accept that it is only the low-energy limit of string or
M-theory?
\item  Are renormalization-group methods a viable way to do non-perturbative
quantum gravity \cite{Reut98,BoRe02}, after the recent discovery of a
non-Gaussian ultraviolet f\/ixed point \cite{LaRe02,ReSa02,LaRe05} of the
renormalization-group f\/low?
\item  Is there truly a singularity avoidance in quantum cosmology
\cite{Espo05a, Espo05b} or string theory \cite{HoSt90, HoMa95, HoMy95}?
\end{enumerate}
We are already facing unprecedented challenges, where the achievements of
spacetime physics and quantum f\/ield theory are called into question.
The years to come will hopefully tell us whether the new mathematical
concepts considered in theoretical physics lead really to a better
understanding of the physical universe and its underlying structures.

\section{Mathematical promises and skepticism --\\ the case of spectral geometry}
\label{s:math}

\subsection*{The base of mathematical skepticism and promises}\label{ss:blabla}
When we write above of ``unprecedented challenges, where the
achievements of spacetime physics and quantum f\/ield theory are
called into question" we are aware that large segments of the
physics community actually are questioning the promised unif\/ied
quantum gravity. We shall not repeat the physicists' skepticism
which was skillfully gathered and elaborated, e.g., by Lee Smolin in~\cite{Smo:TWP}.

Here we shall only add a skeptical mathematical voice, i.e., a
remark made by Yuri Manin in a dif\/ferent context \cite{Man:PAM},
elaborated in \cite{Man:MKI}, and then try to draw a promising
perspective out of Manin's remark.

The Closing round table of the International Congress of Mathematicians (Madrid, August 22--30, 2006)
was devoted to the topic {\em Are pure and applied mathematics drifting apart?}
As panelist, Manin subdivided the mathematization, i.e., the way mathematics can tell us something about the external world,
into three modes of functioning (similarly Bohle, Boo{\ss} and Jensen 1983, \cite{BoBoJe:IEO}, see also \cite{Bo:AIF}):

\begin{enumerate}\itemsep=0pt

\item An {\em (ad-hoc, empirically based) mathematical model} ``describes a certain range of phenomena, qualitatively or quantitatively, but feels uneasy pretending to be something more''. Manin gives two examples for the predictive power of such models, Ptolemy's model of epicycles describing planetary motions of about 150 BCE, and the standard model of around 1960 describing the interaction of elementary particles, besides legions of ad-hoc models which
hide lack of understanding behind a more or less elaborated mathematical formalism of organizing available data.

\item A {\em mathematically formulated theory} is distinguished from an ad-hoc model primarily by its ``higher aspirations. A theory, so to speak, is an aristocratic model." Theoretically substantiated models, such as Newton's mechanics, are not necessarily more precise than ad-hoc models; the coding of experience in the form of a theory, however, allows a more f\/lexible use of the model, since its embedding in a theory universe permits a theoretical check of at least some of its assumptions. A theoretical assessment of the precision and of possible deviations of the model can be based on the underlying theory.

\item A {\em mathematical metaphor} postulates that ``some complex range of
phenomena might be compared to a mathematical construction". As an example, Manin mentions artif\/icial intelligence with its ``very complex systems which are processing information because we have constructed them, and we are trying to compare them with the human brain, which we do not understand very well -- we do not understand almost at all. So at the moment it is a very interesting mathematical metaphor, and what it allows us to do mostly is to sort of cut out our wrong assumptions. If we start comparing them with some very well-known reality, it turns out that they would not work."
\end{enumerate}

Clearly, Manin noted the deceptive formal similarity of the three ways of mathematization which are radically dif\/ferent with respect to their empirical foundation and scientif\/ic status. He expressed concern about the lack of distinction and how that may ``inf\/luence our value systems''. In the words of \cite[p.~73]{Bo:AIF}: ``Well founded applied mathematics generates prestige which is inappropriately generalized to support these quite dif\/ferent applications. The clarity and precision of the mathematical derivations here are in sharp contrast to the uncertainty of the underlying relations assumed. In fact, similarity of the mathematical formalism involved tends to mask the dif\/ferences in the scientif\/ic extra-mathematical status, in the credibility of the conclusions and in appropriate ways of checking assumptions and results \dots Mathematization can~-- and therein lays its success -- make existing rationality transparent; mathematization cannot introduce rationality to a system where it is absent \dots or compensate for a def\/icit of knowledge.''

Asked whether the last 30 years of mathematics' consolidation raise the chance of consolidation also in phenomenologically
and metaphorically expanding sciences, Manin hesitated to use such simplistic terms. He recalled the notion of Kolmogorov
complexity of a piece of information, which is, roughly speaking, ``the length of the shortest programme, which can
be then used to generate this piece of information \dots Classical laws of physics -- such phantastic laws as Newton's law of gravity and Einstein's equations -- are extremely short programmes to generate a lot of descriptions of real physical world situations.
I am not at all sure that Kolmogorov's complexity of data that were uncovered by, say, genetics in the human genome project,
or even modern cosmology data \dots is suf\/f\/iciently small that they can be really grasped by the human mind.''

In spite of our admiration of and sympathy with Manin's
thoughtfulness, the authors of this review shall reverse Manin's
argument and point to the astonishing shortness in the sense of
Kolmogorov complexity of main achievements in one exemplary f\/ield of
mathematics, in spectral geometry to encourage the new unif\/ication
endeavor. Some of the great unif\/ications in physics were preceded by
mature mathematical achievements (like John Bernoulli's unif\/ication
of light and particle movement {\em after} Leibniz' and Newton's
inf\/initesimals and Einstein's general relativity {\em after}
Riemann's and Minkowski's geometries). Other great unif\/ications in
physics were antecedent to comprehensive mathematical theory (like
Maxwell's equations for electro-magnetism {\em long before} Hodge's
and de Rham's vector analysis of dif\/ferential forms). A few great
unif\/ications in physics {\em paralleled} mathematical break-throughs
(like Newton's unif\/ication of Kepler's planetary movement with
Galilei's fall low paralleled calculus and Einstein's 1905 heat
explanation via dif\/fusion paralleled the f\/inal mathematical
understanding of the heat equation via Fourier analysis, Lebesgue
integral and the emerging study of Brownian processes).

In this section, we shall argue for our curiosity about the new
unif\/ication, nourished by the remarkable shortness of basic
achievements of spectral geometry and the surprisingly wide range of
induced (inner-mathematical) explanations.

\subsection*{The power of the index}\label{ss:index}
The most fundamental eigenvalue is the zero, and perhaps the most
fundamental spectral invariant is the index of elliptic problems
over compact manifolds without or with smooth boundary. It is the
dif\/ference between the multiplicity of the zero eigenvalue of the
original  problem and of the adjoint problem. For a
wide introduction to index theory we refer to \cite{BoBl:ASI} and
\cite{BoWo:EBP}, see also the more recent \cite{BoBl:SIO} and
\cite{Espo98}.

In this  subsection we shall summarize the functional analytic framework
of index theory (Fredholm pairs  of subspaces, homotopy invariance
of the index, classifying space), present simple formulations of the
AS and APS theorems,  recall how important geometric invariants can
be written as index of corresponding elliptic problems, emphasize
simple consequences for partitioned manifolds (cutting and pasting
of the index) and delicate consequences for 4D geometry (Donaldson).

The basic object of index theory is a {\em Fredholm pair} $(M,N)\in \cF^2$
of closed subspaces $M,N$ in a f\/ixed Hilbert space $H$, i.e.,
$\dim M\cap N <\infty$, $M+N$ closed in $H$, and $\dim H/(M+N) <\infty$.
We set
\begin{gather*}
\ind (M,N):= \dim M\cap N\, -\, \dim H/(M+N).
\end{gather*}
A bounded operator $F:H\to H$, $F\in \cB$, is Fredholm, $F\in\cF$, if the pair $\bigl(H\times \{0\},
\graph F\bigr)$ is a Fredholm pair in $H\times H$, i.e., if $\dim \ker F<\infty$ and $\dim (H/
\im F) <\infty$ such that $\im F$ is closed and $\ind F = \dim \ker F - \dim \coker F$. Note that
then $\coker F\equiv \ker F^*$,  where $F^*$ denotes the adjoint operator to $F$.

The def\/inition generalizes to closed (not necessarily bounded) operators, typically arising
with elliptic dif\/ferential operators, and to closed relations, typically arising in systems of
dif\/ferential equations when the set of singularities does not have measure zero.

It seems that up to 1950, D. Hilbert's and R. Courant's dictum was generally believed, that
``linear problems of mathematical physics which are correctly posed" satisfy
the so-called {\em Fredholm alternative}, i.e., their index vanishes.
Then G.~Hellwig and I.~Vekua independently showed that the Laplace operator on the disc with
a boundary condition given by a vector f\/ield of winding number $p$ has
index $2-2p$, i.e., $\ne 0$ for $p\ne 1$. A the same time, F.V. Atkinson gave the representation
$\cF\to (\cB/\cK)^\times\to 0$, i.e., by the units of the quotient algebra of the bounded ope\-rators
modulo the ideal $\cK$ of compact operators
(compact operators appear naturally with the inclusion of Sobolev spaces into each other);
a little earlier, J. Dieudonn{\'e} proved the homotopy invariance of the index, more precisely, that $\cF$
decomposes in $\Z$ connected components, distinguished by the index. Later that was generalized by K. J{\"a}nich and
M.F.~Atiyah to a~natural exact sequence of semigroups
\begin{gather}\label{e:jaenich}
[X,\cB^\times]\too[X,\cF]\ftoo{\ind} K(X)\too 0 \qquad\text{ for any compact topological space $X$}.
\end{gather}
Here, $K(X)$ denotes the group of abstract dif\/ference classes of complex vector bundles over~$X$ which was
established by M.F.~Atiyah and F.~Hirzebruch as powerful cohomology theory satisfying Bott periodicity under
suspension of $X$;  $[\cdot,\cdot]$ denotes the semigroup of homotopy classes from one space into another;
and $\ind$ denotes the index bundle which, surprisingly, is well def\/ined even when the dimension of
the kernel of the Fredholm operators parametrized by $X$ varies.
Note that then N.~Kuiper proved that $\cB^\times$ is contractible, i.e., the vanishing of the left term in
\eqref{e:jaenich} yielding an isomorphism between $[X,\cF]$ and $K(X)$. One says that $\cF$ is a {\em classifying space}
for the functor $K$. Setting $X:=\{\operatorname{point}\}$ brings \eqref{e:jaenich} back to Dieudonn{\'e}.

Similar results were obtained by I.M. Singer and M.F.~Atiyah for the topology of the space~$\cF^{\sa}$
of self-adjoint bounded Fredholm operators: it consists of three connected components,
two of them are contractible and the interesting component (those operators which are neither essentially
positive nor essentially negative) is a classifying space
for the functor $K^1$. For $X=S^1$, the isomorphism with $\Z$ is given by the spectral f\/low, see below.

Some of the results can be extended with minor modif\/ications to the spaces $\cC\cR\cF$ of closed Fredholm
relations, $\cC\cF$ (closed
Fredholm operators) or $\cC\cF^{\sa}$ (self-adjoint, not necessarily bounded Fredholm operators).

The interest in the index was nourished by the question (I.M. Singer): ``Why are so many geometric invariants
 integer valued, like the $\wh A$ genus of spin manifolds which is given {\em a priori} as an integral?"
The answer, Singer and Atiyah
found in 1962, see \cite{At:HSE}, is that $\wh A$ and many other important geometric invariants like Euler
characteristic and Thom--Hirzebruch signature can be written as the index of an elliptic dif\/ferential operator.
This is the essence of the celebrated Atiyah--Singer Index Theorem. We give a simplif\/ied version:
Let $M$ be a closed (i.e., compact, without boundary), oriented, smooth Riemannian manifold of dimension $n$,
which is ``trivially'' embedded (i.e., with trivial normal bundle) in the Euclidean space $\R^{n+m}$\,. Let
$E$ and $F$ be Hermitian vector bundles over $M$ and $A$ a linear elliptic operator of order $k\in\Z$,
mapping sections of $E$ to sections of $F$. Its leading symbol $\gs(A)$\footnote{Recall
that for a dif\/ferential operator
$D=\sum\limits_{\ga\in \Z_+^n, \ga_1+\dots+\ga_n\le d} a_\ga(x)\partial_1^{\ga_1}\cdots\partial_n^{\ga_n}$
the leading symbol at the cotangent vector $\xi_i dx^i$ is given by
$\sum\limits_{\ga\in \Z_+^n, \ga_1+\dots+\ga_n = d} a_\ga(x)\xi_1^{\ga_1}\cdots\xi_n^{\ga_n}$.
The leading symbol has an invariant meaning as a section of the bundle $\operatorname{Hom}(E,F)$
over $T^*M$.}
is a bundle isomorphism from~$E$ to~$F$
lifted to the cotangent sphere bundle $S(M)$ and def\/ines an element $[\gs(A)]$ in the abstract ring $K(T^*M)$
of formal dif\/ferences of vector bundles over the full cotangent bundle $T^*M$ of $M$.
Then
\begin{gather}\label{e:as}
\ind A = (-1)^n \gb^{n+m}\bigl([\gs(A)]\boxtimes b^m\bigr),
\end{gather}
where $b\in K(\R^2)$ denotes the Bott class (a generator of $K(\R^2)\simeq \Z$) and
$\gb^{n+m}\! :\! K\bigl(\R^{2(n+m)}\bigr){\to} \Z$ the iteration of the Bott isomorphism.

\eqref{e:as} comprises among numerous other applications
Gauss--Bonnet's expression of the Euler characteristic by a curvature integral,
Hirzebruch's formula of the signature by $L$-polynomials, and the celebrated Riemann--Roch--Hirzebruch Theorem
for complex manifolds.
A special feature of \eqref{e:as} is that we have on the left something which is globally def\/ined, namely by
the multiplicity of the zero-eigenvalues, whereas on the right we have an expression which is def\/ined by
the leading symbol of the operator $A$, i.e., in terms of the coef\/f\/icients of the operator. Actually, the
right hand side can conveniently be written as $\int_M \ga_0(x) \, dx$ where $\ga_0(x)$ is the index density.
It is the constant term in the asymptotic expansion (as $t\to 0$) of the trace dif\/ferences
\begin{gather}\label{e:idensity}
\sum_{\mu\in\spec AA^*}e^{-t\mu}|\gf_\mu(x)|^2 - \sum_{\mu'\in\spec A^*A}e^{-t\mu'}|\gf_{*\mu'}(x)|^2\,,
\end{gather}
where $\mu,\,\gf_\mu$ denote the eigenvalues and eigenfunctions of $AA^*$ and
$\mu',\,\gf_{*\mu'}$ the corresponding objects of $A^*A$.

A simple model of the Index Theorem is provided by the {\em winding number} $\deg(f)$ of a conti\-nuous
mapping $f:S^1\to\C^\times$ of the circle to the punctured plane of non-zero complex numbers. We approximate
$f$ by a dif\/ferentiable $g$ or by a f\/inite Laurent series $h(z)=\sum\limits_{\nu=-k}^{k} a_vz^\nu$ on the disk
$|z|<1$ yielding
\begin{gather*}
\deg(f)=N(h)-P(h)=\frac 1{2\pi \textup{i}}\int \frac{dg}g\,,
\end{gather*}
where $N$ and $P$ denote the number of zeros and poles in $|z|<1$. A direct relation to \eqref{e:as} is that
$\deg(f)=-\ind T_f$\, where $T_f$ denotes the
Toeplitz operator\footnote{In general a Toeplitz operator is the compression of a multiplication operator by a projection,
i.e.\ $PM_fP$. Here $T_f=P\{\text{multiplication by }f\}P$, where $P$ is the orthogonal projection
onto $L^2_+(S^1)$.
}, assigned to $f$ on the half\-space~$L^2_+(S^1)$ spanned
by the functions $z^0,\, z^1,\, z^2,\dots$.

Many quite dif\/ferent proofs of \eqref{e:as} have been given.
At the short end of the scale, regarding the length of proof,
is reducing all cases to (elliptic) Dirac operators and simple
manifolds by cobordism ({\`a} la Hirzebruch) or reducing all cases to the
winding number calculation plus Bott periodicity by embedding in huge Euclidean
space ({\`a la Grothendieck). On the long end (cf.~\cite[p. 281]{AtiBotPat:HEI})
of the scale and promising more insight in the underlying geometry
are the heat equation proofs, inspired by
\eqref{e:idensity} ({\`a} la Minakshisundaram and Pleijel).

To obtain a Fredholm operator from an elliptic operator $A$ over a compact manifold $M$ {\em with boundary} $\gS$
one has to impose suitable boundary conditions. They can be locally def\/ined by bundle homomorphisms or globally
by pseudodif\/ferential projections like the positive spectral ({\em Atiyah--Patodi--Singer}--) projection $P_+$\footnote{i.e.\
the orthogonal projection onto the subspace spanned by the eigenvectors to nonnegative eigenvalues of~$B$.} of the
tangential operator $B$, if $A$ is of Dirac type. Then \eqref{e:as} generalizes to
\begin{gather}\label{e:aps}
\ind A_{P_+} = \int_M \ga_0(x) \, dx - \frac{\dim\ker(B)+\eta(0)}2,
\end{gather}
where $\eta(0)$ denotes the $\eta$ invariant of $B$, see below.
Note that $\dim\ker B$ can jump, e.g., under continuous deformation of the Riemannian metric.
Thus, $\ind A_{P_+ }$ is not homotopy invariant (under homotopies of the metric)
in spite of being the index of a Fredholm operator. Of course this does not
contradict Dieudonn\'e's theorem: $P_+$ does not (necessarily)
depend continuously on the metric and hence so does not the operator $A_{P_+}$.
Consequently, a continuous change of the coef\/f\/icients or the underlying metric structures
can result in a jump of the Fredholm operator from one connected component into another one.

The Atiyah--Patodi--Singer boundary value problem is an example
of a well-posed elliptic boundary value problem. We elaborate a bit on such boundary
value problems for a (total hence formally self-adjoint) Dirac operator $A$. A boundary condition
for $A$ is given by a pseudodif\/ferential operator $P$ of order $0$. It is not a big
loss of generality to assume that $P$ is an idempotent. The domain $\cD(A_P)$ of the
realization $A_P$ then consists of those sections $u$ (of Sobolev class $1$) such that
$P(u_{|\Sigma})=0$. Looking at Green's formula for $A$
\begin{gather}\label{Eq:Green}
    \scalar{Au}{v}-\scalar{u}{Av}=-\int_\Sigma \scalar{\gamma u}{v}d\vol=:\omega(u_{|\Sigma},v_{|\Sigma})
                =\scalar{\gamma u_{|\Sigma}}{v_{|\Sigma}},
\end{gather}
where $\gamma$ is Clif\/ford multiplication by the inward normal vector, we see that the right hand
side of \eqref{Eq:Green} equips $L^2(S_{|\Sigma})$ with a natural Hermitian symplectic structure.
It turns out that a~realization $A_P$ of a well-posed elliptic boundary value problem $P$
is self-adjoint if and only if the range of $P$ is Lagrangian in $(L^2(S_{|\Sigma}),\omega)$.
Well-posedness is originally a microlocal condition. However, for Lagrangian $P$
it is equivalent to the fact that the pair of subspaces $(\ker P,\ran P_+)$ is Fredholm
\cite{BruLes01}.

On partitioned manifolds $M=M_-\cup_{\gS} M_+$ one has the {\em Bojarski Formula}
\begin{gather}\label{e:boj}
\ind A = \ind(H_-(A),H_+(A)),
\end{gather}
where $(H_-(A),H_+(A))$ denotes the Fredholm pair of Cauchy data spaces
(=boundary values of solutions to the equation $Au=0$ on $M_+$ respectively $M_-$)
along the separating hypersurface $\gS$.
Contrary to the precise Novikov additivity of Euler characteristic and signature
under cutting and pasting, decomposition formulas for all other indices contain
non-vanishing corrections in terms of the geometry of $\Sigma$ and the gluing
dif\/feomorphisms.

Novikov additivity of Euler characteristic and signature can be
explained by the index theorem, but, being combinatorial, is more
easily obtained by purely topological arguments. Some deep insight
in the geometry of 3- and 4-dimensional dif\/ferentiable manifolds,
however, seems to {\em rely} on the index theorem. Most famous is S.
Donaldson's theorem \cite{DonKro90}
that any closed smooth simply-connected
4-manifold $M$ with positive-def\/inite intersection form $s_M$ can
be written as the boundary of a $5$-manifold, is a connected sum of
complex projective spaces and $s_M$ is trivial, i.e., in standard
diagonal form. The proof depends on the investigation of the moduli
space of solutions of the Yang--Mills equation of gauge-theoretic
physics. Note that by a famous theorem of M. Freedman any unimodular
(integral) quadratic form can appear as $s_M$ for exactly one or two
simply-connected oriented {\em topological} 4-manifolds. By that
Freedman conf\/irmed Poincar{\'e}'s Conjecture in dimension 4.
Moreover, according to Freedman there exist $10^{51}$ mutually non
homeomorphic oriented simply-connected topological manifolds with
40-dimensional 2-homology (i.e., with 40 bubbles), but, according
to Donaldson, only one (more precisely two) permit a dif\/ferentiable
structure. Note that all piecewise linear 4-manifolds can be
equipped with a dif\/ferentiable structure, i.e., the dif\/ference
between dif\/ferentiable and solely topological cannot be perceived by
appealing to piecewise linear constructions.

As another consequence of Donaldson's theorem, $\R^4$ admits an exotic dif\/ferentiable structure which renders it
not dif\/feomorphic to $\R^4$ with its usual dif\/ferentiable structure.

\subsection*{Other spectral invariants}\label{ss:specinv}
While the index, e.g., of the chiral Dirac operator $A^+$ only measures the (chiral) asymmetry
of the zero eigenspaces, there are three other spectral invariants which can provide much more
information about a geometry one may be interested in: the spectral f\/low, the $\eta$ invariant
and the determinant.

The {\em spectral flow} $\SF$ measures the
net sign change of the eigenvalues around zero for a curve in $\cC\cF^{\sa}$ (cf.\ the previous section).
For f\/ixed endpoints, it does not change under a homotopy of the curve. For bounded
self-adjoint (and neither essentially positive nor essentially negative) Fredholm operators, the spectral f\/low
yields an isomorphism of the fundamental group on the integers.
Only recently \cite{Joa03} it was shown that the same statement also holds for $\cC\cF^{\sa}$
equipped with the graph (=gap) topology. Surprisingly, the latter space
is connected as was observed f\/irst in \cite{BooLesPhi01}.

Referring to \eqref{e:boj} (which is for dif\/ferential operators only meaningful on
even-dimensional manifolds), we consider a curve of (total)
Dirac operators over a partitioned
manifold. Then the Cauchy data spaces along the separating hypersurface $\gS$ are Fredholm pairs of
Lagrangian subspaces of the symplectic Hilbert space $L^2(\gS,S|_{\gS})$ (cf.\ \eqref{Eq:Green}) and we have
\begin{gather}\label{e:boj-yn}
\SF\{A_s\}_{0\le s\le 1} = \Mas\{(H_-(A_s),H_+(A_s))\}_{0\le s\le 1}\,,
\end{gather}
where $\Mas$ denotes the Maslov index, which adds the (f\/inite and signed) intersection dimensions. Note that
the left side of \eqref{e:boj-yn} is def\/ined by the spectrum (actually, a generalization of the
Morse index), while the right side is more ``classical" and, in general, easier to calculate (actually, a~generalization of
the number of conjugate points). Ongoing research shows that
formula \eqref{e:boj-yn} can be obtained in much greater generality.

Note also that we have $\ind (\dd s + B_s) = \SF \{B_s\}$ for any curve of elliptic operators of f\/irst order
over a closed manifold $\gS$ with unitarily equivalent ends, when the operator on the left side is induced
on the corresponding mapping torus.

It turns out, that not only the chiral and the dynamic asymmetry of the
zero eigenspaces can be measured,
but also the asymmetry of the total spectrum of an operator $A$ of Dirac
type on a closed manifold. Thus,
roughly speaking, we have a formally self-adjoint
operator with a~real discrete spectrum which is nicely spaced without f\/inite
accumulation points and with an inf\/inite number of eigenvalues on both sides
of the real line. In close analogy with the def\/inition of the zeta function
for essentially positive elliptic operators like the Laplacian, we set
\begin{gather}\label{e:eta-def}
\eta(z):=\sum_{\lambda\in\operatorname{spec}(A)\setminus\left\{  0\right\}  }
\operatorname*{sign}(\lambda)\,\lambda^{-z}
 = \frac 1{\gG(\frac{z+1}2)}\int_{0}^{\infty}t^{\frac{z-1}{2}
}\Tr\big(Ae^{-tA^{2}}\big)\,dt.
\end{gather}
Clearly, the formal sum $\eta(z)$ is well def\/ined for complex
$z$ with $\Re(z)$ suf\/f\/iciently large, and it vanishes for a symmetric spectrum.

For comparison we give the corresponding formula for the zeta function of the
(positive) Dirac Laplacian $A^{2}$:
\begin{gather}
\zeta_{A^{2}}(z):=\Tr((A^{2})^{-z})=\frac
{1}{\Gamma(z)}\,\int_{0}^{\infty}t^{z-1}\Tr\big(e^{-t{A}^{2}}\big)\,dt.
\label{e:zeta-def}
\end{gather}
For the zeta function, we must assume that $A$ has no vanishing
eigenvalues (i.e., $A^2$ is positive). Otherwise the integral on
the right side is divergent. (The situation, however, can be cured by
subtracting the orthogonal projection onto the kernel of $A^{2}$
from the heat operator before taking the trace.) For the eta function, on the
contrary, it clearly does not matter whether there are 0-eigenvalues and
whether the summation is over all or only over the nonvanishing eigenvalues.

The derivation of \eqref{e:eta-def} and \eqref{e:zeta-def} is completely
elementary for $\Re(z)>\frac{1+\dim M}{2}$, resp.\ $\Re(z)>\frac{\dim
M}{2}$, where $M$ denotes the underlying manifold.
Next, recall the heat trace expansion of elliptic operators
which holds in great generality:
\begin{gather}\label{asymexp}
\Tr(B e^{-t T})\sim_{t\to 0+} \sum_{j=0}^\infty a_j(T,B) t^{(j-\dim M-b)/2}+
    \sum_{j=0}^\infty (b_j(T,B)+c_j(T,B)\log t) t^j,
\end{gather}
$b=\text{order of } B$, for any (pseudo)dif\/ferential operator $B$, and any
positive def\/inite self-adjoint elliptic (pseudo)dif\/ferential operator $T$ (here for
simplicity of order $2$) \cite{GruSee95}.

Using \eqref{asymexp}, it follows at once that $\eta(z)$ (and $\zeta(z)$)
admit a meromorphic extension to the
whole complex plane. However, a priori $0$ is a pole. In general the residue
at $0$ (which is a multiple of the coef\/f\/icient of $\log t$ in \eqref{asymexp})
 of the \emph{generalized $\zeta$-function}
$\Tr(B T^{-z})=\frac{1}{\Gamma(z)}\int_0^\infty t^{z-1}\Tr(Be^{-tT}) dt $
turns out to be an invariant of $B$ independently of $T$: it is the
celebrated non-commutative residue \cite{Eliz98, Eliz99, Eliz00, Byts04}
discovered independently by V.~Guillemin~\cite{Gui85} and M. Wodzicki
\cite{Wod87}. The residue of $\Id$ is easily seen to vanish.
Consequently the $\zeta$-func\-tion~\eqref{e:zeta-def} is always regular at $0$.
In fact Wodzicki proved that the residue of any (pseudo)dif\/fe\-ren\-tial idempotent vanishes
and from this one can conclude that also the $\eta$-func\-tion~\eqref{e:eta-def}
is always regular at $0$.

In the decade or so following 1975, it was generally believed that the
existence of a f\/inite eta invariant was a very special feature of operators of
Dirac type on those closed manifolds which are boundaries. Later Gilkey \cite{Gil81}
proved that the eta-function is regular at $0$ for any self-adjoint elliptic operator on
a closed manifold, cf.\ also Branson and Gilkey \cite{BraGil92}. Only after the seminal paper
by {Douglas} and {Wojciechowski} \cite{DoWo91} it was gradually realized that
globally elliptic self-adjoint boundary value problems for operators of Dirac
type also have a f\/inite eta invariant. Today, there are quite dif\/ferent
approaches to obtain that result.

Historically, the eta invariant appeared for the f\/irst time in the 1970s as an
error term showing up in the index formula for the APS spectral boundary value
problem of a Dirac operator $A$ on a compact manifold $M$ with
smooth boundary $\Sigma$ (see above \eqref{e:aps}). More
precisely, what arose was the eta invariant of the tangential operator (i.e.,
the induced Dirac operator over the closed manifold $\Sigma$). Even in that
case it was hard to establish the existence and f\/initeness of the eta invariant.

The eta-invariant and the spectral f\/low are intimately related: let $A_s$ be
a smooth family of Dirac operators on a closed manifold. Then
$\xi(A_s):=\frac 12(\dim\ker A_s+\eta(A_s))$ has only integer jumps and hence the
\emph{reduced eta-invariant} $\widetilde\eta(A_s):=(\frac 12(\dim\ker A_s+\eta(A_s))
\mod \Z)\in \R/\Z$ varies smoothly. The variation of the reduced eta-invariant
is known to be local:
\begin{gather*}
     \frac{d}{ds}\widetilde \eta(A_s)=-\frac{1}{\sqrt{\pi}}
           \operatorname{Pf}_{t\to 0+}\sqrt{t}\Tr\big(\dot A_s e^{-tA_s^2}\big)
\end{gather*}
where $\operatorname{Pf}$ (partie f\/inie) is a short hand for the coef\/f\/icient of $t^0$ in the asymptotic
expansion. Because of the integer jumps of $\xi(A_s)$ the fundamental theorem of calculus now reads
\begin{gather*}
     \eta(A_1)-\eta(A_0)=2 \SF\{A_s\} + 2\int_0^1 \frac{d}{ds} \widetilde\eta(A_s) ds.
\end{gather*}
It is a matter of convention how to count $0$-modes of $A_0$, $A_1$ for the spectral f\/low and we ignore it here.

Unlike the index and spectral f\/low on closed manifolds, we have neither an
established functional analytical nor a topological frame for discussing
$\eta(0)$, nor is it given by an integral of a~locally def\/ined
expression. On the circle, e.g., consider the operator
\begin{gather*}
\mathcal{D}_{a}:=-\textup{i}\frac{d}{dx}+a
=e^{-\textup{i}xa}\mathcal{D}_{0}e^{\textup{i}xa},
\end{gather*}
so that $\mathcal{D}_{a}$ and $\mathcal{D}_{0}$ are locally
unitarily equivalent, but not globally. In
 particular, $\eta_{\mathcal{D}_{a}}(0)=-2a$ depends on $a$.

The reduced $\eta$-invariant depends, however, only on f\/initely many terms of the
complete symbol of $A$.


By the Cobordism Theorem, the index vanishes on any closed (odd-dimensional)
manifold~$\Sigma$ for any elliptic operator which is the chiral tangential
operator of a Dirac type operator on a suitable (even-dimensional) manifold
which has $\Sigma$ as its boundary. This can be explained by the vanishing of
the induced symplectic form over $\gS$. Correspondingly, we have in even
dimensions, that the eta invariant vanishes on any closed manifold $\Sigma$ for
any elliptic operator which is the tangential operator of a Dirac type operator
on a suitable manifold which has $\Sigma$ as its boundary. This can be
explained by the induced precise symmetry of the spectrum due to the
anti-commutativity of the tangential operator with Clif\/ford multiplication.

To us, the eta invariant of a Dirac type operator can be best understood as the phase
of the zeta regularized determinant. But what is the {\em determinant} of an operator with
an inf\/inite number of eigenvalues?

From the point of view of functional analysis, the only natural
concept is the {\em Fredholm determinant} of operators of the form
$e^\alpha$ or, more
generally, $\Id +\ga$  where $\ga$ is of trace class. We recall the formulas
\begin{gather*}
{\det}_{\Fr} \ e^{\ga} = e^{\Tr \ga} \qquad \tand\qquad
{\det}_{\Fr} (\Id + \ga) = \sum_{k=0}^{\infty}\Tr \wedge^k\ga .
\end{gather*}
The Fredholm determinant is notable for obeying the product rule, unlike other generalizations of the determinant to inf\/inite
dimensions where the error of the product rule leads to new
invariants.

Clearly, the parametrix of a Dirac operator
leads to operators for which the Fredholm determinant can be def\/ined,
but the relevant information about the spectrum of the Dirac operator
does not seem suf\/f\/iciently maintained. Note also that Quillen and
Segal's construction of the {\it determinant line bundle} is based on
the concept of the Fredholm determinant, though without leading to a
{\it number} when the bundle is non-trivial.

Another concept is the $\zeta$-{\it function regularized
determinant}, based on Ray and Singer's observation that, formally,
\[
\det T = \prod \lambda_j = \exp\left\{\sum \ln \lambda_j e^{-z\ln
\lambda_j}|_{z=0}\right\}
= e^{-\frac d{dz}\zeta_T(z)|_{z=0}},
\]
where $\gz_T(z):=\sum\limits_{j=1}^\infty \gl_j^{-z} =
\frac 1{\gG(z)}\int_0^\infty t^{z-1}\, \Tr e^{-tT}\,dt  $ as in \eqref{e:zeta-def}.
As explained before, for a {\em positive definite self-adjoint
elliptic} operator $T$ of
second order, acting on sections of a
Hermitian vector bundle over a closed manifold $M$,
the function $\zeta_T(z)$ is holomorphic for $\Re(z)$
suf\/f\/iciently large and can be extended meromorphically to the whole
complex plane with $z=0$ no pole.

The previous def\/inition does not apply immediately to the
Dirac operator $A$ which has inf\/initely many positive eigenvalues ${\lambda_j}$
and negative eigenvalues ${-\mu_j}$.
As an example, consider again the operator $\mathcal{D}_{a}$
with $\spec \mathcal{D}_{a}=\{k+a\}_{k\in\Z}$.

On choosing the branch $(-1)^{-z} = e^{\textup{i}{\pi}z}$, we f\/ind
\begin{gather*}
\gz_{A}(z) =
\sum \lambda_j^{-z} + \sum (-1)^{-z}\mu_j^{-z}
=\12\left\{\zeta_{A^2}\left(\frac z2\right) + \eta_{A}(z)\right\} +
\12 e^{\textup{i}{\pi}z}
\left\{\zeta_{A^2}\left(\frac z2\right) -\eta_{A}(z)\right\},
\end{gather*}
where
$\eta_{A}(z)$ is def\/ined as in \eqref{e:eta-def}. Thus:
\begin{gather*}
\gz'_{A}(0) = \12\gz'_{A^2}(0) +
\frac{\textup{i}\pi}2\left\{\gz_{A^2}(0) - \eta_{A}(0)\right\}
\qquad \tand\qquad {\det}_\gz A = e^{-\gz'_{A}(0)}.
\end{gather*}
This relation between the phase of the $\zeta$-determinant and the $\eta$-invariant
does in fact hold in great generality, at least for self-adjoint elliptic operators on
closed manifolds.

There are a few classes of operators for which the $\zeta$-determinant can be calculated
quite explicitly. The most satisfying theory is available for one-dimensional operators.
Consider a~one-dimensional Schr\"odinger operator on the interval $[0,1]$ with,
to make things a bit interesting, a~\emph{regular singular} potential
\begin{gather*}
    L:=-\frac{d^2}{d x^2}+V(x),\qquad V(x)=\frac{q(x)}{x^2(1-x)^2},
\end{gather*}
with some smooth function $q(x)$, $0\le x\le 1$, $q(0),q(1)\ge -1/4$. The last condition
guarantees that $L$ is bounded below and thus we can consider the \emph{Friedrichs extension}
$L^{\cF}$. Physically this means that the potential walls at $0$, $1$ are inf\/initely high
and hence the quantum-mechanical particle is trapped in the interval $[0,1]$.
Then the $\zeta$-determinant of $L^{\cF}$ can be expressed in terms of the fundamental
solutions of the homogeneous dif\/ferential equation $Lf=0$ (see \cite{Les98}, also for a history of the
problem). Indeed, choose solutions
$\varphi$, $\psi$ of this equation satisfying $\varphi(x)\sim_{x\to 0} x^{\nu_0+1/2}$,
$\psi(x)\sim_{x\to 0} (1-x)^{\nu_1+1/2}$, $\nu_j=\sqrt{a(j)+1/4}$, $j=0,1$. Then
\begin{gather*}
   \detz(L^{\cF})=\frac{\pi (\psi \varphi'-\varphi \psi')}{2^{\nu_0+\nu_1}\Gamma(\nu_0+1)\Gamma(\nu_1+1)}.
\end{gather*}
Note that the \emph{Wronskian}  $\psi \varphi'-\varphi \psi'$ is a constant.

This formula can to a certain extent be generalized to higher dimensions. Note that the solutions
of the homogeneous dif\/ferential equation are determined by their Cauchy data at the boundary.
Therefore the Wronskian should be viewed as a Fredholm determinant\footnote{Here it is a f\/inite-dimensional determinant, since the boundary
is $0$-dimensional and hence the space of sections on the boundary is
f\/inite-dimensional.} on the boundary.

For Dirac operators,  S.~Scott and K.P.~Wojciechowski \cite{ScWo00} found
an exciting relation between the Fredholm determinant and the
zeta-regularized determinant:
Let $A$ be a
Dirac operator over a compact manifold $M$ with boundary
$\Sigma$ and tangential operator $\begin{pmatrix} 0&B^-\\B^+&0\end{pmatrix}$.
The decomposition of the tangential operator is induced by the $\pm {\rm i}$-eigenbundle
decomposition of the natural symplectic structure (cf.~\eqref{Eq:Green}); in the odd-dimensional case
this is the chiral decomposition on the even-dimensional boundary.
Let $P$ be the APS or another suitable pseudodif\/ferential projection
def\/ining a regular elliptic boundary value problem for $A$.
Then the Cauchy data space~$H_+(A)$ (i.e., the range of the Calder{\'o}n projection $C_+$)
and the range of $P$
can be written as the graphs of unitary, elliptic operators of order 0, $K$,
resp. $T$ which dif\/fer from the operator $(B^{+}B^{-})^{-1/2} B^{+}:
C^{\infty}(\Sigma;S^{+}|_{\Sigma}) \to C^{\infty}(\Sigma;S^{-}|_{\Sigma})$ by
a smoothing operator. Then
\begin{gather*}
{\det}_{\zeta}A_{P} = {\det}_{\zeta}A_{C_+%
} \,\cdot\ {\det}_{\operatorname{Fr}} \tfrac12 (\operatorname{I} +
KT^{-1}).
\end{gather*}
This formula shows how the $\zeta$-determinant depends on the choice of the boundary
condition. As in the one-dimensional case, the space of solutions of the homogeneous
operator equation $Au=0$, through its Calder\'on projector, enters crucially.

Finally we mention another class of operators where relative $\zeta$-determinants
can be calculated: let $M$ be a closed oriented surface with a Riemannian metric $g$.
Let $h$ be a smooth function on $M$ and consider the conformally changed metric $g_h=e^{2h}g$.
By $\Delta_h$ we denote the (positive def\/inite) Laplacian with respect to the metric $g_h$.
Then the \emph{Polyakov formula} reads
\begin{gather}\label{Polyakov}
  \log\frac{\detz \Delta_h}{\detz \Delta_0}=-\frac{1}{12\pi} \int_M |\nabla h|^2 \dvol_0
    -\frac{1}{6\pi} \int_M K_0 h \dvol_0 +\log\frac{\vol (M,g_0)}{\vol (M,g_h)}.
\end{gather}
Here $\nabla$ is the Levi-Civita connection and $K_0$ is the Gaussian curvature of the metric $g_0$.
The formula \eqref{Polyakov} goes back to Polyakov \cite{Pol81-1,Pol81-2}. It
was used in the work of Osgood, Phillips, and Sarnak \cite{OsgPhiSar88-2}
who showed that exactly the constant curvature metrics are the extremals
of the determinant in a conformal class. More interestingly, they did this \emph{without} using the Uniformization
Theorem. Rather their method reproves this theorem.

Formulas similar to \eqref{Polyakov} were obtained by T.~Branson and B.~{\O}rsted
for the conformal Laplacian in four dimensions \cite{BraOrs92}.

Finally we would like to provide a glimpse at pasting formulas for $\eta$-invariants
and $\zeta$-determinants since in the last decade substantial progress has been made.
Compared to pasting formulas for the index (cf.~\eqref{e:boj}, Novikov additivity)
this is technically much more involved. The case which is most similar to the Euler characteristic
is that of the analytic torsion. Using the celebrated Cheeger--M\"uller theorem on the relation between
analytic and combinatorial torsion one obtains pasting formulas more or less by employing the doubling
trick: on the double $\widetilde M=M\bigcup_{\partial M} M$ dif\/ferential forms split naturally
into a sum of forms obeying the relative resp. absolute boundary conditions on $\partial M$
\cite{Luc93}. A direct analytic approach is also possible but technically much more involved
\cite{Vis95}. It was Vishik's great idea to consider on a partitioned manifold $M_1\cup_\Sigma M_2$
the following family of boundary conditions for dif\/ferential forms
\begin{gather*}
   \sin\Theta i_1^* \omega= \cos \Theta i_2^* \omega,
\end{gather*}
where $i_j$ is the pullback to $\Sigma$ from $M_j$. This family interpolates between the
direct sum of the absolute boundary condition on $M_1$ and the relative boundary condition
on $M_2$ on the one hand ($\Theta=0$) and the continuous transmission condition on $\Sigma$ ($\Theta=\pi/4$) on the other hand.

This motive was later adapted to Dirac operators and the $\eta$-invariant \cite{BruLes99}.
The f\/inal gluing formula for the $\eta$-invariant of a Dirac operator on a partitioned manifold
reads
\begin{gather*}
    \eta(D)-\eta(D_P|M_2)-\eta(D_{I-P}|M_1)=\operatorname{Mas}(P_1,P,I-P_2).
\end{gather*}
Here, $P$ is a well-posed Lagrangian boundary condition on  $M_1$ as explained after \eqref{Eq:Green}
and $\operatorname{Mas}(P_1,P,I-P_2)$ is a Maslov type triple index involving the boundary condition
$P$ and the Calder{\'o}n projectors (projectors on the Cauchy data spaces) on $M_1$, $M_2$
\cite{KirLes04}.

In a titanic ef\/fort Park and Wojciechowski \cite{ParWoj02,ParWoj06}
eventually succeeded to prove an analytic surgery formula for the
$\zeta$-determinant. We refrain from giving a precise formulation,
see loc.~cit.

\subsection*{Spectral triples and other new ideas and results of noncommutative geometry}
\label{ss:noncom}

Noncommutative Geometry is an area of mathematics which has been dominated
by the work of Alain Connes over the last 20--25 years.
The basic idea is that instead of point sets (e.g.~manifolds) one studies the
coordinate ring of (smooth) functions.
This point of view has been around in algebraic geometry for decades but it was
Connes who showed that also manifolds and index
theory can be understood from this perspective.

The starting point of Noncommutative \emph{Topology} is the well-known
Gelfand-representation
theorem for commutative $C^*$-algebras: to any compact space $X$ we
can associate its ring $C(X)$ of continuous functions.
This is a commutative $C^*$-algebra. Conversely,
if $\mathcal A$ is a commutative $C^*$-algebra then we can f\/ind $X$ as the
\emph{spectrum}
of $\mathcal A$, i.e., the space of maximal ideals.
$X\mapsto C(X)$ and the Gelfand functor
$\mathcal A\mapsto \operatorname{spec} \mathcal A$ are in fact mutually
inverse category equivalences.
Thus, from this perspective the space $X$ and the coordinate ring (ring
of ``position operators'') contain the same information.

It was Connes' fantastic discovery that this correspondence between space and algebra
can be pushed much further in the dif\/ferentiable category. The basic object is now
a (locally convex) algebra $\cA$ (playing the role of $C^\infty(M))$. Indispensable basic
tools on dif\/ferentiable manifolds are dif\/ferential forms and de Rham cohomology.
The natural replacement for dif\/ferential forms is the Hochschild cohomology
$H^n(\cA)$ which is the cohomology of the complex $(C^n(\cA),b)$ where $C^n(\cA)$
are the (continuous) linear functionals on the $(n+1)$-fold tensor product
$\cA\otimes\cdots\otimes\cA$ and $b:C^n(\cA)\to C^{n+1}(\cA)$
is the well-known Hochschild boundary map
\begin{gather*}
  b \phi  (a_0 , \ldots , a_{n+1}) =
  \sum_{j=0}^n (-1)^j   \phi  ( a_0 , \ldots , a_j a_{j+1} ,
  \ldots , a_{n+1} )
+ (-1)^{n+1} \phi ( a_{n+1} a_0 , a_1 , \ldots , a_n ).
\end{gather*}
Restricting $b$
to the \emph{cyclic} cochains $C^n_\lambda(\cA)$\footnote{$\varphi(a_n,a_0,\ldots,a_{n-1})=(-1)^n \varphi(a_0,\ldots,a_n)$}
yields another complex whose
cohomology is the cyclic cohomology $HC^n(\cA)$.
The celebrated Hochschild--Kostant--Rosenberg--Connes Theorem
says that for a compact smooth manifold the map which sends
a de Rham current $C$ (i.e., an element in the dual space of
dif\/ferential forms) to the linear form
\[
       \varphi_C(f_0,\ldots,f_n)
:=\langle C,f_0df_1\wedge\cdots\wedge df_n\rangle
\]
induces f\/irstly an isomorphism between de Rham currents and the Hochschild
cohomology $H^\bullet(C^\infty(M))$ and secondly an isomorphism between ($\Z_2$-graded)
de Rham homology and the (periodic) cyclic cohomology of the algebra $C^\infty(M)$.
Under this correspondence the exterior derivative (on currents) is sent to (a multiple of)
Connes' operator $B:C^n(\cA)\to C^{n-1}(\cA)$,
\begin{gather*}
  B \phi (a_0 , \ldots ,  a_{n-1})  =
  \sum_{j=0}^{n-1} (-1)^{(n-1)j}  \phi ( 1 , a_j , \ldots , a_{n-1},
  a_0 , \ldots ,  a_{n-1} )
 \nonumber\\
\phantom{B \phi (a_0 , \ldots ,  a_{n-1})  = }{} -
  \sum_{j=0}^{n-1} (-1)^{(n-1)j}  \phi (a_j , 1 , a_{j+1} , \ldots , a_n ,
  a_0 , \ldots , a_{j-1} ) .
\end{gather*}
$B$ is important for a double complex realization of cyclic cohomology.
In fact $Bb=-bB$, $B^2=0$, $b^2=0$ and hence one obtains
a double complex with entries $C^{n,m}(\cA):=C^{n-m}(\cA)$ and the total
cohomology of this double complex is the periodic cyclic cohomology.

The story is getting even more interesting when it comes to index theory. Given a
closed (even-dimensional) spin manifold with its Dirac operator $D$, the Dirac operator determines
a~\emph{spectral triple} $(\cA=C^\infty(M),L^2(S),D)$ where $L^2(S)$ is the Hilbert space
of $L^2$-sections of the spinor bundle (the space of spinor f\/ields). The algebra
$\cA$ obviously acts by multiplication as bounded operators on $L^2(S)$, and
we have that $[D,a]$ is bounded for all $a\in\cA$.

It should be noted f\/irst that the spectral triple $(\cA,L^2(S),D)$ is as good as the
Riemannian manifold $(M,g)$: f\/irst the geodesic distance can easily be reconstructed
from the spectral triple, i.e. \cite[p.~544]{Con94}
\begin{gather*}
  d(p,q)=\sup\bigsetdef{|a(p)-a(q)|}{a\in\cA, \|[D,a]\|\le 1}.
\end{gather*}
Even more, Connes' Spin Manifold Theorem  \cite[Chapter~11]{GraVarFig01} in particular implies that
the Riemannian metric may be recovered from the spectral triple. If one assumes more structure on
the spectral triple (axioms of a noncommutative spin geometry) then even the manifold may
be reconstructed from a noncommutative spin geometry with commutative algebra $\cA$.
The list of axioms is a bit lengthy, though. The f\/irst attempt
\cite{Ren01} to prove that Connes' original list of axioms suf\/f\/ices fell short of its goal
as was observed by the reviewer of \cite{Ren01}.
Quite recently, the result was proved for a slightly stronger set of axioms \cite{Ren06}.

An abstract spectral triple $(\cA,H,D)$\footnote{$\cA$ an algebra represented as bounded operators on the Hilbert
space $H$, $D$ an unbounded self-adjoint operator in $H$, $[D,a]$ bounded for $a\in\cA$.} is
therefore the natural noncommutative model of a smooth Riemannian manifold.
Spectral triples come in two f\/lavors, even and odd. Not surprisingly, the spectral
triple of a spin manifold is even/odd if the dimension of the manifold is even/odd.
To explain the basic index theory in the noncommutative world let us restrict ourselves to the
even case. An even spectral triple comes with a chirality operator $\gamma$
which anticommutes with $D$. For any idempotent $e$ (vector bundle,
projective module) in the matrix algebra $M_k(\cA)$ we then obtain a Fredholm operator
$eD^+e:e(H^+)^k\longrightarrow e(H^-)^k$. Its index is given by the pairing between the
K-homology class of $(\cA,H,D)$ and the K-theory class of $e$. Taking Chern characters
(in (entire) periodic cyclic (co)homology) we have
\begin{gather}\label{Eq:indexpairing}
    \ind(eD^+e)=\langle \Ch^\bullet(D),\ch_\bullet(e)\rangle.
\end{gather}
Here,
\begin{gather*}
  \ch_\bullet(e) := \tr_0(e)+ \sum_{j=1}^\infty (-1)^j \frac{(2j)!}{j!}
  \tr_{2j} \Big( \left(e - \tfrac{1}{2} \right)\otimes e^{\otimes (2j)}\Big),
\end{gather*}
is the Chern character in K-theory\footnote{$\tr_0:M_n(\cA)\to A$ is the generalized trace map, $\tr_{2j}$ is its obvious
extension to tensor products.}. $\Ch^\bullet(D)$ is the JLO-Chern character, i.e.,
\begin{gather*}
       \Ch^n(D)(a_0,\ldots,a_k)
    =\int_{\Delta_n} \tr(\gamma a_0 e^{-\sigma_0 D^2}[D,a_1]e^{-\sigma_1 D^2}\cdots [D,a_n] e^{-\sigma_n D^2}) d\sigma,
\end{gather*}
where $\Delta_n:=\bigsetdef{(\sigma_0,\ldots,\sigma_n)\in\R^{n+1}}{\sigma_j\ge 0, \sigma_0+\dots+\sigma_n=1}$
is the standard $n$-simplex.

Replacing $D$ by $\sqrt{t} D$, $t>0$, does not change the left and right
hand sides of
\eqref{Eq:indexpairing}. Under the assumption that we do have short-time
asymptotic expansions
like \eqref{asymexp} in our spectral triple\footnote{The proper notion is that of a spectral triple with discrete dimension spectrum.}
it was shown in \cite{ConMos95} that also the JLO-cocycle
$\Ch^n(\sqrt{t}D)$ has short time asymptotic expansions.

The transgression formula
\begin{gather*}
   -\frac{d}{dt} \Ch^n(tD)=b \slch^{n-1}(tD,D)+B\slch^{n+1}(tD,D)
\end{gather*}
(the expression for $\slch$ is explicitly known) then shows that the JLO-cocycle may be replaced
by a cohomologous cocycle consisting of ``local data'' which still computes the index
\eqref{Eq:indexpairing}. The details of this are the content of the celebrated Local Index Theorem in
noncommutative geometry of Connes and Moscovici \cite{ConMos95}.


\section{Noncommutative geometry\\ and
the standard model with gravity}

The full action of the standard model of particle physics is quite involved and
spelled out it f\/ills about a page \cite[Section 4.1]{ChaConMar06}.
It has been known
for a long time that bits of the standard model Lagrangian are
intimately related
to the heat trace expansion \eqref{asymexp}. To be more specif\/ic, let $D$ be
a Dirac type operator on a closed manifold.
Recall that locally a Dirac operator has the form
$g^{ij}c(\partial_i)\partial_j+ \cdots$,
i.e.\ its leading symbol at a (co)tangent vector $v$ is given by
Clif\/ford multiplication by $v$.
Consequently its square is an operator of Laplace type which locally
takes the form $D^2=g^{ij}\partial_i\partial_j+\cdots$.

The heat trace expansion \eqref{asymexp} takes the form
\begin{gather*}
\Tr (e^{-t D^2})\sim_{t\to 0+} \sum_{j=0}^\infty
A_{j/2}(D^2) t^{(j-\dim M)/2},
\end{gather*}
where $A_{j/2}(D^2)=\int_M a_{j/2}(x;D^2)d\vol_M(x)$ is the integral of a local
expression $a_{j/2}(x;D^2)$ in the metric, the f\/ibre metric,
and the coef\/f\/icients
of $D$. From the very construction it is immediately clear that the functional
$A_{j/2}(D^2)$ is invariant under the natural action of the
dif\/feomorphism group.
Such functionals are not easy to f\/ind from scratch and the complexity
of the expression for $A_{j/2}$ increases fast. For \emph{differential}
operators $A_{j/2}=0$ for $j$ odd.
For a Dirac operator on a~Clif\/ford bundle $E$, the f\/irst few
$A_j's$ are known explicitly. We mention
\begin{gather*}
a_0\big(x, D^2\big)=(4\pi)^{-\dim M/2} \rank (E),
\qquad A_0\big(D^2\big)=(4\pi)^{-\dim M/2} \rank(E) \vol(M),\nonumber\\
a_1\big(x, D^2\big)=(4\pi)^{-\dim M/2} \tr \left(-\frac{R}{6}+\mathcal E\right),
\end{gather*}
where $R$ denotes the scalar curvature and
$\mathcal E=\nabla^*\nabla-D^2$ is the $0$th order
term in the Lichnerowicz formula for $D$. Hence, for the spin
Dirac operator $A_1(D^2)$ is nothing but the Einstein action.

The appearance of the Einstein action in the heat trace expansion may lead to
the following speculation: can one cook up a Dirac operator in such a way
that the full action of the standard model appears in the short-time
(i.e. high energy expansion) of the heat trace? The answer to this question
is probably no. However, the stunning result of Chamseddine, Connes and
Marcolli~\cite{ChaConMar06} is that the answer is yes if we consider
the Dirac operator on a mildly noncommutative space.

To explain this, we f\/irst summarize a few relations between the heat trace,
the $\zeta$-function and the non-commutative residue. Again, let $D$
be a Dirac type operator on the closed manifold~$M$. For a pseudodif\/ferential
operator $A$ the noncommutative residue is def\/ined by
\begin{gather*}
\regint A:=\Res_{s=0}\Tr(A |D|^{-s})\\
\phantom{\regint A \ }{}=-2 \text{ times the coef\/f\/icient of }
\log t \text{ in the asymptotic expansion of } \Tr\big(A e^{-tD^2}\big).\nonumber
\end{gather*}
The noncommutative residue is indeed independent of the choice of $D$.
For the Dirac operator itself there is an obvious relation between
the noncommutative residue of $|D|^{-k}$ and the heat coef\/f\/icients. Namely,
for a non-negative integer $k$
\begin{gather*}
\regint |D|^{-k}=\Res_{s=k} \zeta_{|D|}(s)=\frac{2 A_{(\dim M -k)/2}(D^2)}{\Gamma(k/2)}
\end{gather*}
(with the understanding $1/\Gamma(0)=0$).

Using simple properties of the Mellin transform, one therefore derives
the following result (cf.~\cite[Appendix]{ChaConMar06}): if
$f$ is an even smooth rapidly decaying function on the real line then
one has the following asymptotic expansion:
\begin{gather}
\Tr(f(D/\Lambda))\sim_{\Lambda\to\infty}
f(0) A_{\frac 12 \dim M}(D^2)\Lambda^0+\sum_{k=1}^{\dim M} f_k
\regint |D|^{-k}\, \Lambda^k+o(1)\nonumber\\
\phantom{\Tr(f(D/\Lambda))}{}\sim_{\Lambda\to\infty} f(0) A_{\frac 12 \dim M} (D^2)\Lambda^0
+\sum_{k=1}^{\dim M} f_k \frac{2 A_{(\dim M -k)/2}(D^2)}{\Gamma(k/2)}\,\Lambda^k+o(1),\label{asymspecac}
\end{gather}
where $f_k=\int_0^\infty u^{k-1}f(u)du$. This asymptotic expansion is the key
for the understanding of the high energy expansion of the
\emph{spectral action}.

Some decade ago A.~Chamseddine and A.~Connes \cite{ChaCon97}
proposed the spectral action
\begin{gather}\label{specaction}
   \Tr(f(D/\Lambda)+\tfrac 12 \langle J\psi,D\psi\rangle
\end{gather}
on a noncommutative space $(\cA,\cH,D)$. The stunning relation to the
standard model is as follows.

Let $M$ be a closed $4$-dimensional spin manifold. Furthermore, let
$F$ be a f\/inite spectral triple over the algebra
$\cA=\C\oplus \mathbb{H}\oplus M_3(\C)$.
The algebra $\cA$ is deliberately chosen such that the natural group
of local gauge transformations of the space $M\times F$ is
the semidirect product of
$C^\infty(M,\operatorname{U}(1)\times \operatorname{SU}(2)
\times\operatorname{SU(3)})$ by the dif\/feomorphism group of $M$.

The spin Dirac operator on $M$ and the Dirac operator on the f\/inite
spectral triple $(\cA,H,D)$ (this is after all a self-adjoint endomorphism
of a f\/inite-dimensional Hilbert space) give rise to a spectral
triple $(C^\infty(M,\cA),\cH,D)$ over the noncommutative algebra
$C^\infty(M,\cA)$. The Dirac operator on the f\/inite spectral triple
$(\cA,H,D)$ depends on $31$ real parameters, which physically correspond to
the masses for leptons and quarks etc. The main result
of \cite{ChaConMar06} states that the asymptotic formula
$\Lambda\to\infty$ for the spectral action functional \eqref{specaction}
yields the full Lagrangian of the standard model with neutrino mixing
and Majorana mass terms. Note that through the spin manifold $M$
gravity is naturally built into the model.

Let us look at \eqref{asymspecac}. Then the result of \cite{ChaConMar06}
says that for the noncommutative spectral triple sketched above
the Lagrangian of the standard model shows up in the coef\/f\/icients
of the expansion \eqref{asymspecac}. This is a far reaching
generalization of the above observation that the Einstein action
is the $A_1$-term in the expansion for the spin Dirac operator.

\section{Heat-kernel asymptotics and quantum gravity}\label{Section6}

In classical and quantum f\/ield theory, as well as in the current
attempts to develop a quantum theory of the universe and of
gravitational interactions, it remains very useful to describe
physical phenomena in terms of dif\/ferential equations for the
variables of the theory, supplemented by boundary conditions
for the solutions of such equations. For example, the problems
of electrostatics, the analysis of waveguides, the theory of
vibrating membranes, the Casimir ef\/fect, van der Waals forces,
and the problem of how the universe could evolve from an initial
state, all need a careful assignment of boundary conditions.
In the latter case, if one follows a~functional-integral approach, one
faces two formidable tasks: (i)~the specif\/ication of the geometries
occurring in the ``sum over histories" and matching the assigned
boundary data; (ii)~the choice of boundary conditions on metric
perturbations which may lead to the evaluation of the one-loop
semiclassical approximation.

Indeed, while the full functional integral for quantum gravity is a
fascinating idea but remains a formal tool, the one-loop
calculation may be put on solid ground, and appears particularly
interesting because it yields the f\/irst quantum corrections
to the underlying classical theory (although it is well known that
quantum gravity based on Einstein's theory is not perturbatively
renormalizable). Within this framework, it is of crucial importance
to evaluate the one-loop divergences of the theory under
consideration. Moreover, the task of the theoretical physicist is
to understand the deeper general structure of such divergences.
For this purpose, one has to pay attention to all geometric
invariants of the problem, in a way made clear by a branch of
mathematics known as invariance theory. The key geometric
elements of our problem are hence as follows.

A Riemannian geometry $(M,g)$ is given, where the manifold $M$
is compact and has a boun\-dary~$\partial M$ with induced metric $\gamma$, and
the metrics $g$ and $\gamma$ are positive-def\/inite. A vector
bundle over~$M$, say $V$, is given, and one has also to consider
a vector bundle $\widetilde V$ over $\partial M$. An operator of
Laplace type, say $P$, is a second-order elliptic operator with
leading symbol given by the metric (more precisely, it has scalar leading
symbol $g^{ab}k_{a}k_{b}$). Thus, one deals with a map
from the space of smooth sections of $V$ onto itself,
\begin{gather*}
P: C^{\infty}(M,V) \rightarrow C^{\infty}(M,V),
\end{gather*}
which can be expressed in the form
\begin{gather*}
P=-g^{ab} \; \nabla_{a}^{V} \; \nabla_{b}^{V}-E,
\end{gather*}
where $\nabla^{V}$ is the connection on $V$, and $E$ is an
endomorphism of $V$: $E \in {\text {End}}(V)$. Moreover, the
boundary operator is a map
\begin{gather*}
{\cal B}: C^{\infty}(M,V) \rightarrow C^{\infty}({\partial
M},{\widetilde V}),
\end{gather*}
and contains all the informations on the boundary conditions
of the problem. Since gauge theories need a generalization of
Robin boundary conditions, we consider a boundary operator
of the form (the operation of restriction to the boundary
being implicitly understood)
\begin{gather}\label{68}
{\cal B}=\nabla_{N}+{1\over 2}\Bigr[\Gamma^{i}
{\widehat \nabla}_{i}+{\widehat \nabla}_{i}\Gamma^{i}\Bigr]+S .
\end{gather}
With our notation, $\nabla_{N}$ is the normal derivative
operator $\nabla_{N} \equiv N^{a}\nabla_{a}$ ($N^{a}$ being the
inward-pointing normal to $\partial M$), $S$ is an endomorphism
of the vector bundle $\widetilde V$, $\Gamma^{i}$ are
endomorphism-valued vector f\/ields on $\partial M$,
and ${\widehat \nabla}_{i}$
denotes tangential covariant dif\/ferentiation with respect to the
connection induced on $\partial M$. More precisely, when sections
of bundles built from $V$ are involved, ${\widehat \nabla}_{i}$ means
\[
\nabla_{\partial M}^{({\rm lc})} \otimes 1
+ 1 \otimes \nabla ,
\]
where $\nabla_{\partial M}^{({\rm lc})}$ denotes the
Levi-Civita connection of the boundary of $M$. Hereafter, we
assume that $1+\Gamma^{2} >0$, to ensure strong ellipticity
of the boundary value problem (see def\/inition on pp.~69--70 of
\cite{Espo98}).

The case of mixed boundary conditions corresponds to the
possibility of splitting the bund\-le~$V$, in a neighbourhood
of $\partial M$, as the direct sum of two bundles, say $V_{1}$
and $V_{2}$, for each of which a~boundary operator of the
Dirichlet or (generalized) Robin type is also given. The former
involves a projection operator, say $\Pi$, while the latter may
also involve the complementary projector, $1 - \Pi$, and the
metric of $V$, say $H$:
\begin{gather*}
{\cal B}_{1}=\Pi ,
\qquad
{\cal B}_{2}=(1 - \Pi)\Bigr[H \nabla_{N}
+\tfrac{1}{2}\bigr(\Gamma^{i}{\widehat \nabla}_{i}
+{\widehat \nabla}_{i}\Gamma^{i}\bigr)+S \Bigr].
\end{gather*}
We can now come back to our original problem, where only the
boundary operator \eqref{68} occurs, and investigate its ef\/fect on
heat-kernel asymptotics. Indeed, given the heat
equation for the operator $P$, its kernel, i.e. the heat kernel,
is, by def\/inition, a solution for $t > 0$ of the equation
\begin{gather*}
\left({\partial \over \partial t}+P \right)U(x,x';t)=0,
\end{gather*}
jointly with the initial condition
\begin{gather*}
\lim_{t \to 0}\int_{M}dx' \sqrt{{\text {det}}g(x')} \;
U(x,x';t) \rho(x')=\rho(x) ,
\end{gather*}
and the boundary condition
\begin{gather*}
\bigl[{\cal B}U(x,x';t)\bigr]_{\partial M}=0.
\end{gather*}
The f\/ibre trace of the heat-kernel diagonal, i.e.\
${\text {Tr}}U(x,x;t)$, admits an asymptotic expansion which
describes the {\it local} asymptotics, and involves interior
invariants and boundary invariants. Interior invariants are
built universally and polynomially from the metric, the
Riemann curvature~$R_{\; \; bcd}^{a}$ of~$M$, the bundle
curvature, say $\Omega_{ab}$, the endomorphism $E$, and their
covariant derivatives. By virtue of Weyl's work on the invariants
of the orthogonal group, these polynomials can be found by using
only tensor products and contraction of tensor arguments.
The asymptotic expansion of the integral
\begin{gather*}
\int_{M}dx \sqrt{{\text {det}}\,g} \; {\text {Tr}}U(x,x;t) \equiv
{\text {Tr}}_{L^{2}} \bigl(e^{-tP}\bigr),
\end{gather*}
yields instead the {\it global} asymptotics. For our purposes,
it is more convenient to weight $e^{-tP}$ with a smooth function
$f \in C^{\infty}(M)$, and then consider the asymptotic expansion
\begin{gather}\label{75}
{\text {Tr}}_{L^{2}}\bigl(f e^{-tP} \bigr)  \equiv
\int_{M}dx \sqrt{{\text {det}}g} \; f(x){\text {Tr}} U(x,x;t)
\sim (4\pi t)^{-m/2} \sum_{l=0}^{\infty}
t^{l/2}A_{l/2}(f,P,{\cal B}).
\end{gather}
Hereafter, $m$ is the dimension
of $M$, and the coef\/f\/icient $A_{l/2}(f,P,{\cal B})$ consists
of an interior part, say $C_{l/2}(f,P)$, and a boundary part,
say $B_{l/2}(f,P,{\cal B})$. The interior part vanishes for
all odd values of $l$, whereas the boundary part only vanishes
if $l=0$. The interior part is obtained by integrating over $M$ the
linear combination of local invariants of the appropriate
dimension mentioned above, where the coef\/f\/icients of the linear
combination are {\it universal constants}, independent of $m$.
Moreover, the boundary part is obtained upon integration over
$\partial M$ of another linear combination of local invariants.
In that case, however, the structure group is $O(m-1)$,
and the coef\/f\/icients of linear combination are {\it universal
functions}, independent of $m$, unaf\/fected by conformal
rescalings of $g$, and invariant in form (i.e.\ they are
functions of position on the boundary, whose form is independent
of the boundary being curved or totally geodesic). It is thus
clear that the general form of the $A_{l/2}$ coef\/f\/icient is a
well posed problem in invariance theory, where one has to take
all possible local invariants built from $f$, $R_{\; \; bcd}^{a}$,
$\Omega_{ab}$, $K_{ij}$, $E$, $S$, $\Gamma^{i}$ and their covariant
derivatives (hereafter, $K_{ij}$ is the extrinsic-curvature
tensor of the boundary), eventually integrating their linear
combinations over $M$ and $\partial M$. For example, in the
boundary part $B_{l/2}(f,P,{\cal B})$, the local invariants
integrated over $\partial M$ are of dimension $l-1$ in
tensors of the same dimension of the second fundamental form
of the boundary, for all $l \geq 1$. The universal functions
associated to all such invariants can be found by using the
conformal-variation method described, for example,
in~\cite{Espo98}, jointly
with the analysis of simple examples and particular cases.

In other words, recurrence relations of algebraic nature exist
among all universal functions, and one can therefore use the
solutions of simple problems to determine completely the
remaining set of universal functions for a given value of the
integer $l$ in the asymptotic expansion \eqref{75}.
The detailed investigation of the coef\/f\/icients $A_{1}$, $A_{3/2}$
and $A_{2}$ when the boundary operator is given by equation~\eqref{68} and
all curvature terms are non-vanishing is performed in~\cite{Avra98, Dowk99}
One then f\/inds the result (which holds for all integer values
of $l \geq 2$)
\begin{gather}\label{76}
A_{l/2}(f,P,{\cal B})={\widetilde A}_{l/2}(f,P,{\cal B})
+\int_{\partial M}{\text {Tr}}\bigl[a_{l/2}(f,R,\Omega,K,E,
\Gamma,S)\bigr],
\end{gather}
where ${\widetilde A}_{l/2}(f,P,{\cal B})$ is formally analogous
to the purely Robin case, but replacing the universal constants
in the boundary terms with universal functions, whereas
$a_{l/2}$ is a linear combination of all local invariants of the
given dimension which involve contractions with $\Gamma^{i}$. Our
task is now to derive an algorithm for the general form of
$a_{l/2}$, since it helps a lot to have a formula that clarif\/ies
the general features of a scheme where the number of new invariants
is rapidly growing. Indeed, from~\cite{Avra98, Dowk99},
we know that, in $a_{1}$,
only one new invariant occurs: $fK_{ij}\Gamma^{i}\Gamma^{j}$,
whereas in~$a_{3/2}$ 11 new invariants occur, obtained by
contraction of $\Gamma^{i}$ with terms like (tensor indices are
here omitted for simplicity)
\[
fK^{2}, fKS, f{\widehat \nabla}K,f{\widehat \nabla}S,fR,
f\Omega,f_{;N}K.
\]
In $a_{2}$, the number of new invariants is 68: 57 involve
contractions of $\Gamma^{i}$ with terms like
\begin{gather*}
fK^{3},fK^{2}S,fKS^{2},fRK,f\Omega K, fEK, fRS, f \Omega S,
fK{\widehat \nabla}K, fS {\widehat \nabla}K,
\\
fK{\widehat \nabla}S,fS{\widehat \nabla}S,
f{\widehat \nabla}{\widehat \nabla}K,
f{\widehat \nabla}{\widehat \nabla}S,
f \nabla R, f \nabla \Omega, f \nabla E,
\end{gather*}
10 local invariants involve contractions of $\Gamma^{i}$ with
contributions like
\[
f_{;N}K^{2},f_{;N}KS,f_{;N}{\widehat \nabla}K,
f_{;N}{\widehat \nabla}S,f_{;N}R, f_{;N} \Omega,
\]
and the last invariant is $f_{;NN}K_{ij}\Gamma^{i}\Gamma^{j}$.
It is thus clear that the knowledge of all local invariants
in $a_{l/2}$ plays a role in the form of $a_{(l+1)/2}$, and
one can write the formulas
\begin{gather*}
a_{1}=f\sum_{i=1}^{i_{1}}{\cal U}_{i}^{(1,1)} \,
I_{i}^{(1)},
\\
a_{3/2}=f\sum_{i=1}^{i_{2}}{\cal U}_{i}^{(3/2,3/2)} \,
I_{i}^{(3/2)}+f_{;N}\sum_{i=1}^{i_{1}}
{\cal U}_{i}^{(3/2,1)} \, I_{i}^{(1)},
\\
a_{2}=f\sum_{i=1}^{i_{3}}{\cal U}_{i}^{(2,2)} \, I_{i}^{(2)}
+f_{;N}\sum_{i=1}^{i_{2}}
{\cal U}_{i}^{(2,3/2)} \; I_{i}^{(3/2)}
+f_{;NN}\sum_{i=1}^{i_{1}}{\cal U}_{i}^{(2,1)} \,
I_{i}^{(1)}.
\end{gather*}
With our notation, $i_{1}=1$, $i_{2}=10$, $i_{3}=57$, and
${\cal U}_{i}^{(x,y)}$ are the universal functions, where
$i$ is an integer $\geq 1$, $x$ is always equal to the order
$l/2$ of $a_{l/2}$, and $y$ is equal to the label of the
invariant~$I_{i}^{(y)}$, which does not contain $f$ or
derivatives of $f$ and is of dimension $2y-1$ in $K$ or in
tensors of the same dimension of $K$.

These remarks make it possible to write down a formula which
holds for all $l \geq 2$:
\begin{gather}\label{80}
a_{l/2}(f,R,\Omega,K,E,\Gamma,S)=\sum_{r=0}^{l-2}f^{(r)}
\sum_{i=1}^{i_{l-r-1}}{\cal U}_{i}^{(l/2,(l-r)/2)}[\Gamma^{2}]
I_{i}^{(l-r)/2}[R,\Omega,K,E,\Gamma,S],
\end{gather}
where $f^{(r)}$ is the normal derivative of $f$ of order $r$
(with $f^{(0)}=f$), and square brackets are used for the
arguments of universal functions and local invariants,
respectively. The equations~\eqref{76} and \eqref{80} represent the
desired parametrization of heat-kernel coef\/f\/icients with
genera\-lized boundary conditions, provided that the $\Gamma^{i}$
are covariantly constant.

One has now to evaluate the universal functions in the
general formulas for $A_{3/2}$, $A_{2}$, $A_{5/2}$
and so on. For the coef\/f\/icients $A_{3/2}$
and $A_{2}$, results of a limited nature are available in~\cite{Avra98, Dowk99}, which show that all universal functions are generated
from $\sqrt{1+\Gamma^{2}}$ and ${1\over \sqrt{-\Gamma^{2}}}
\text {Artanh} \sqrt{-\Gamma^{2}}$.
Upon completion of this hard piece of work, one
could perform the evaluation of all universal functions for
$A_{5/2}(f,P,{\cal B})$ as well, possibly with the help of
computers. For this purpose, one has to combine
the conformal-variation method with the analysis
of simpler cases. One then obtains a
quicker and more elegant derivation of the coef\/f\/icient
$A_{1}(f,P,{\cal B})$. There are thus reasons to expect that,
in the near future, all heat-kernel coef\/f\/icients with
generalized boundary conditions may be obtained via a computer
algorithm in a relatively short time. This adds evidence in
favour of the understanding of general mathematical structures
being very helpful in providing the complete solution of
dif\/f\/icult problems in physics and mathematics. In particular,
from the point of view of quantum f\/ield theory in curved
manifolds, this would mean {\it an entirely geometric understanding
of the f\/irst set of quantum corrections to the underlying
classical theory}, with the help of invariance theory,
functorial methods and computer programs. This is the case because,
on the one hand, heat-kernel coef\/f\/icients $A_{l/2}(f,P,{\cal B})$ in
the asymptotic expansion \eqref{75} are obtained from a geometric
construction as we said, while, on the other hand, the coef\/f\/icient
$A_{l/2}$ yields the one-loop divergence of the corresponding theory
in $l$-dimensional space-time. Such one-loop divergences are indeed
the f\/irst set of quantum corrections to the classical theory. This has
implications for the old unif\/ication program of Section~\ref{Section2}, because it
shows the perturbative limits of its quantization: one-loop quantum gravity
based on the Einstein--Hilbert action is only on-shell f\/inite, i.e.\ upon
imposing the vacuum Einstein equations
\[
R_{ab}-\tfrac{1}{2}g_{ab}R=0 \quad \Longrightarrow \quad R_{ab}=0.
\]
The reason is that only three geometric invariants can be built in such a
case, i.e.
\[
R_{abcd}R^{abcd},\qquad R_{ab}R^{ab},\qquad R^{2},
\]
bearing in mind that the fourth one, given by the wave operator acting
on the scalar curvature, has vanishing integral if spacetime has empty
boundary. The above invariants should be integrated over the spacetime
manifold, so that the integral representation of the Euler number makes it
possible to deal eventually with two invariants only: $R_{ab}R^{ab}$
and $R^{2}$. But both of them vanish if and only if the metric solves the
vacuum Einstein equations in four dimensions.

In Euclidean quantum gravity, if one uses the de Donder
gauge-f\/ixing functional, one f\/inds a~boundary operator formally
analogous to the one in equation~\eqref{68} but imcompatible with the strong
ellipticity of the boundary value problem \cite{Avra99}. Only recently,
in~\cite{Espo05a, Espo05b}, has one found a viable way out in
the particular case of the Euclidean 4-ball, and more work is in order
on this key issue, which has also implications for singularity avoidance
in quantum cosmology, as we mentioned in Section~\ref{Section3}. This framework is
relevant also for the new unif\/ication outlined in Section~\ref{Section2}, as is
shown in~\cite{Barv07}. In this work, the authors study a toy model
of brane-induced gravity for the calculation of its one-loop ef\/fective
action, and obtain the inverse-mass asymptotic expansion of the ef\/fective
action and its ultraviolet divergences, which are found to be non-vanishing
in all spacetime dimensions. They also obtain the heat-kernel asymptotics
associated with generalized boundary conditions containing tangential
derivatives (cf.\ equation~\eqref{68}). In addition to the usual powers of the
$t$-parameter in the expansion \eqref{75}, they f\/ind also logarithmic terms
or powers multiple of one quarter. Such a property is considered in the
context of strong ellipticity of the boundary value problem, which can be
violated under certain conditions~\cite{Barv07}.

\appendix
\section[Infinite-dimensional manifolds in quantum gravity]{Inf\/inite-dimensional manifolds in quantum gravity}
We would like to stress here that inf\/inite-dimensional manifolds are the
natural arena for studying the quantization of the gravitational f\/ield,
even prior to considering a space-of-histories formulation. In~\cite{Gero71}, an approach is proposed where the states and operators
emerge as certain scalar and vector f\/ields on an inf\/inite-dimensional
manifold ${\cal G}$ of classical solutions of Einstein's equations.
Such a formalism essentially reproduces quantum electrodynamics when
${\cal G}$ is replaced by the space of square-integrable solutions of
Maxwell's equations. According to the author of~\cite{Gero71}, a
quantum theory of gravitation is obtained if one can solve two problems:
\renewcommand{\labelenumi}{\textup{\arabic{enumi}.}}
\begin{enumerate}
\itemsep=0pt

\item Impose a manifold structure on a suitable collection of solutions
of the Einstein equations (e.g.~without sources).
\item Determine tensor f\/ields $T_{a}^{\; b}$ and $G^{ab}$ on that manifold,
subject to
\begin{gather*}
T_{a}^{\; b} \; T_{b}^{\; c}=-\delta_{a}^{\; c},
\qquad
G^{ab}=G^{ba}, \qquad  T_{c}^{\; a} G^{cb}=-T_{c}^{\; b} G^{ca}.
\end{gather*}
\end{enumerate}

The solution of these two problems for any classical system of interacting
bosonic f\/ields should produce a quantum theory. In general relativity, one
might select for ${\cal G}$ the collection of all asymptotically simple
solutions of the Einstein equations, the main steps being as follows~\cite{Gero71}. Let ${\overline M}= M \cup {\cal I}$ be a
manifold with boundary, where the boundary ${\cal I}$ consists of two
disjoint copies of $S^{2} \times \R$, and the interior of $M$ is
$\R^{4}$. The main idea is to incorporate into ${\overline M}$ all
structure which is independent of the metric chosen. One therefore requires
that a conformal factor $\Omega$ is given on ${\overline M}$ such that
\begin{gather*}
\Omega >0 \  \ {\rm on} \  \ M, \qquad
\Omega=0 \ \ {\rm on} \  \ {\cal I},
\end{gather*}
while the gradient of $\Omega$ is nonvanishing on ${\cal I}$, whose conformal
structure is required to be given. Let now ${\cal G}$  be the collection of
all source-free solutions of the Einstein equations on $M$ which are
asymptotically simple with conformal factor $\Omega$ and conformal
inf\/inity ${\cal I}$, and which reproduce the given conformal structure
on ${\cal I}$. A vector in ${\cal G}$ at a point $g \in {\cal G}$ (the
point $g$ being a solution of the Einstein equations) is a linearized
solution $\gamma_{\alpha}^{\; \beta}$ which is regular on ${\cal I}$. Thus,
the desired tensor f\/ield $T_{a}^{\; b}$ is a linear mapping on linearized
solutions at $g$ which, applied twice to $\gamma_{\alpha}^{\; \beta}$,
gives $-\gamma_{\alpha}^{\; \beta}$. One could obtain such a mapping by
expressing the linearized solution in terms of its initial data on past
null inf\/inity, and then altering the phase of the data in a suitable way.
Last, but not least, the desired metric $G_{ab}$ on ${\cal G}$ should
assign a real number to each linearized solution. One might use for this
purpose the integral \cite{Gero71}
\begin{gather*}
I \equiv \int_{M} \gamma_{\alpha}^{\; \beta}
\; \gamma_{\beta}^{\; \alpha} dV.
\end{gather*}

To sum up, there are at least three sources of inf\/inite-dimensionality
in quantum gravity:
\begin{enumerate}
\itemsep=0pt
\item  The inf\/inite-dimensional Lie group (or pseudo-group) of spacetime
dif\/feomorphisms, which is the invariance group of general relativity
in the f\/irst place \cite{DeWi65,Stre07}.
\item The inf\/inite-dimensional space of histories in a functional-integral
quantization \cite{DeWi03, DeWi05}.
\item The inf\/inite-dimensional Geroch space of asymptotically simple
spacetimes \cite{Gero71}.
\end{enumerate}

\subsection*{Acknowledgments}
G. Esposito is grateful to the INFN for f\/inancial
support, and to IMFUFA (Roskilde University), where part of this
material was f\/irst presented.

This paper is based on joint discussions among the authors when preparing the
Summer School 2008 on ``Quantum Gravity'', see \url{http://QuantumGravity.ruc.dk}.

\pdfbookmark[1]{References}{ref}
 \LastPageEnding

\end{document}